\documentclass[sn-nature]{sn-jnl}
\usepackage[table,cmyk]{xcolor}
\usepackage{amsmath,amsthm,,amsfonts,amssymb,commath,braket,booktabs}
\usepackage{calc,graphicx,bm,tikz,float}
\usepackage{xr}
\usepackage{multirow}
\usepackage{hyperref}
\definecolor{oeawblue}{cmyk}{0.9,0.68,0,0}
\definecolor{iqoqiblue}{cmyk}{0.76,0.11,0,0}
\definecolor{iffsred}{cmyk}{0.12,0.94,0.87,0.34}
\definecolor{thpurple}{cmyk}{0.65,1.0,0.0,0.2}
\definecolor{uestcblue}{cmyk}{0.99,0.78,0.16,0.03}

\usepackage{dcolumn}
\usepackage{braket}
\usepackage{xcolor}
\usepackage{tabularx}
\usepackage{mathtools}
\usepackage{pifont}
\usepackage{subcaption}

\usepackage{caption}
\newcommand\blamb{\bm{\lambda}}
\newcommand\tr{{\rm tr}}
\newcommand{\comment}[1]{{}}

\newtheorem{theorem}{Theorem}
\newtheorem{lemma}{Lemma}

\hypersetup{
  pdftitle={Error-Mitigated Quantum Metrology via Enhanced Virtual Purification},
  pdfauthor={Lin et al.},
  pdfstartview=Fit,
  pdfpagelayout=SinglePage,
  colorlinks,
  linkcolor=uestcblue,
  citecolor=uestcblue,
  urlcolor=thpurple}
\usepackage[figure]{hypcap}

\usepackage[percent]{overpic}

\begin{document}

\title{Error-Mitigated Quantum Metrology via Enhanced Virtual Purification}

\author[1,4]{\fnm{Xiaodie} \sur{Lin}}
\author*[1,2,3]{\fnm{Haidong} \sur{Yuan}}\email{hdyuan@mae.cuhk.edu.hk}

\affil[1]{\orgdiv{Department of Mechanical and Automation Engineering}, \orgname{The Chinese University of Hong Kong}, \orgaddress{\city{Hong Kong SAR}, \country{China}}}

\affil[2]{\orgdiv{The Hong Kong Institute of Quantum Information Science and Technology}, \orgname{The Chinese University of Hong Kong}, \orgaddress{\city{Hong Kong SAR}, \country{China}}}

\affil[3]{\orgdiv{State Key Laboratory of Quantum Information Technologies and Materials}, \orgname{The Chinese University of Hong Kong}, \orgaddress{\city{Hong Kong SAR}, \country{China}}}

\affil[4]{\orgdiv{College of Computer and Data Science}, \orgname{ Fuzhou University}, \orgaddress{\city{Fuzhou}, \postcode{350116}, \country{China}}}

\abstract{
Quantum metrology stands as a leading application of quantum science and technology, yet noise often constrains its precision and sensitivity. In near-term quantum metrology, existing protocols largely depend on virtual state purification, but significant noise accumulation and additional noise from the implementations of these protocols can impede their effectiveness. We propose enhanced virtual channel purification to address these problems, yielding enhanced virtual state purification as a by-product. Within sequential quantum metrology schemes, our error analysis reveals substantial bias reduction and quantum advantages in sampling cost when the number of encoding channels is ${\mathcal{O}}(p^{-1})$, where $p$ represents the error rate of encoding channels. In this range, our methods demonstrate significant improvements in parameter estimation precision and robustness against practical noise, as evidenced by numerical simulations for both single- and multi-parameter tasks. Particularly, these methods can naturally extend beyond quantum metrology, indicating their broad applicability in quantum information and quantum computation.
}

\maketitle

\section*{Introduction}

Quantum metrology is one of the most promising applications of quantum science and technology~\cite{PhysRevLett.96.010401,maccone2021,RevModPhys.89.035002,RevModPhys.90.035005}. Quantum resources, such as coherence and entanglement, can be employed to enhance the sensitivity to reach the Heisenberg limit. However, due to the existence of decoherence or other sources of noise, the achievable precision and sensitivity of quantum metrology could be limited~\cite{Davidovich2011,Madalin2012,Dobrzanski18}.

Quantum error correction (QEC) is a promising approach to address noise limitations in quantum metrology, with numerous studies demonstrating its potential to enhance measurement precision~\cite{PhysRevLett.112.150802,PhysRevLett.112.080801,robust2015,PhysRevLett.115.200501,PhysRevLett.122.040502,Liang2018}. {Nevertheless, certain scenarios exist where QEC cannot effectively suppress noise, e.g., noise is proportional to the signal, as established by the Hamiltonian-not-in-Lindblad-span (HNLS) condition~\cite{zhou18}. Furthermore, practical implementation of QEC faces significant challenges, including noise in ancillary qubits, imperfect error correction operations, large timescales of error correction schemes, and often unrealistic theoretical assumptions~\cite{Shettell_2021}. Therefore, quantum error mitigation (QEM)~\cite{RevModPhys.95.045005,PhysRevLett.119.180509,PhysRevX.8.031027,PhysRevResearch.3.033098,PhysRevX.11.041036} has emerged as a practical alternative to mitigate errors in quantum metrology within the current stage of quantum technology~\cite{PhysRevLett.129.250503,PhysRevA.109.022410,kwon2025virtualpurificationcomplementsquantum,hama2023quantum}. Unlike QEC, which aims to correct errors in individual quantum circuits, QEM significantly relaxes the experimental precision requirements by post-processing outputs from an ensemble of noisy circuits. Particularly,  in scenarios where the HNLS condition is violated, QEM can also be employed to complement QEC strategies for quantum metrology~\cite{kwon2025virtualpurificationcomplementsquantum}.} 

Since virtual state purification (VSP)~\cite{PhysRevX.11.041036,PhysRevX.11.031057} does not require characterization of noise models, existing error-mitigated quantum metrology protocols are based primarily on it~\cite{PhysRevLett.129.250503,PhysRevA.109.022410,kwon2025virtualpurificationcomplementsquantum}. The basic idea of VSP is to exponentially suppress errors using collective measurements of $m$ copies of the target quantum state $\rho$ to measure the values with respect to the state $\rho^m/\tr(\rho^m)$. However, this approach imposes stringent requirements: (i) both input and ideal output states must remain pure throughout the protocol, and (ii) the noiseless component must dominate the noisy output state. These conditions become particularly challenging in quantum metrology applications, where repeated channel use for parameter encoding leads to large system sizes or deep quantum circuits. Consequently, error accumulation causes significant misalignment between the dominant noisy component and the ideal output state, leading to the failure of VSP.

To address these limitations, we construct an error-mitigated quantum metrology protocol based on virtual channel purification (VCP)~\cite{liuVirtualChannelPurification2024}. VCP offers distinct advantages over VSP by allowing error mitigation at arbitrary circuit locations rather than solely at the final output state. Therefore, errors in quantum metrology can be suppressed before they are too large to mitigate. However, both VSP and VCP rely on performing collective quantum operations on different quantum subsystems to ``purify" the target quantum state. These operations may introduce substantial noise to physical experiments, thereby limiting the practical effectiveness of these methods.

To further address this problem, we propose enhanced virtual channel purification that incorporates probabilistic error cancellation (PEC)~\cite{PhysRevLett.119.180509,PhysRevX.8.031027} specifically tailored for VCP, referred to as VCP-PEC. While the combination of PEC with VCP has been discussed previously~\cite{liuVirtualChannelPurification2024}, our analysis of the impact of noise at various positions within VCP circuits reveals the existence of noise at certain positions that introduces no systematic errors. Consequently, PEC is strategically applied only at critical positions to efficiently cancel errors. The strategy for applying PEC can be naturally adapted for VSP, leading to the enhanced virtual state purification (VSP-PEC). Here, we focus on the sequential feedback scheme of quantum metrology, as it has been shown to outperform the parallel scheme in Hamiltonian parameter estimation and is more experiment-friendly~\cite{PhysRevLett.117.160801}.

Notice that the cost of exactly mitigating general noise grows exponentially with the number of noisy quantum operations~\cite{PhysRevLett.131.210601,PhysRevLett.131.210602,endo22}, which can easily overwhelm the polynomially enhanced precision in quantum metrology~\cite{PhysRevA.88.022318,PhysRevLett.124.190503,PhysRevLett.123.210502,PhysRevLett.127.200504}. Notably, the error analysis of VCP-PEC indicates the significant reduction of bias while maintaining quantum advantage when the number $N$ of utilized encoding channels, each with an error rate $p$, is no more than the order of $p^{-1}$.  For $N$ in the range of ${\mathcal{O}}(p^{-1})$, the effectiveness of both enhanced virtual purification methods is systematically demonstrated through numerical simulations for both single- and multi-parameter estimation tasks. Our results not only exhibit a significant improvement in the precision of the estimated parameters but also showcase their robustness against practical noise and imperfect noise model characterization. In addition, we emphasize that the above analysis is applicable beyond quantum metrology, indicating the versatile potential of our enhanced virtual purification methods across a wide spectrum of quantum information processing and quantum computation applications.

\section*{Methods}
\subsection*{Settings and typical quantum error mitigation methods}

\begin{figure}[!ht]
  \centering
  \includegraphics[width=0.98\columnwidth]{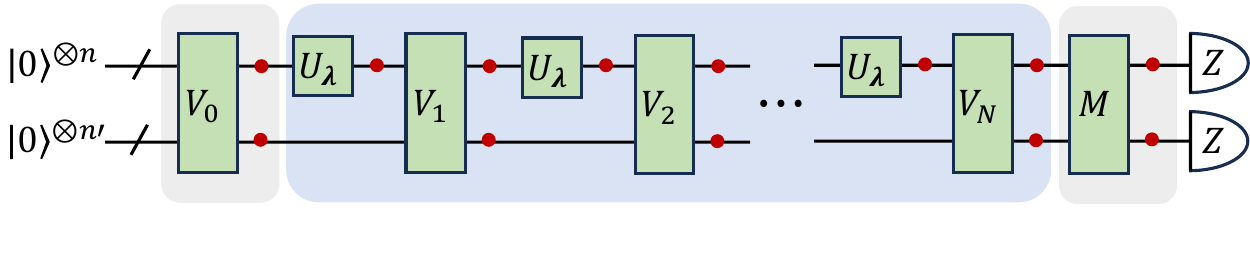}
  \caption{Sequential feedback scheme of quantum metrology.  Green boxes represent quantum gates, while red circles indicate local noise occurring immediately after each quantum gate. Particularly, the first gray box, the blue box and the final gray box represent the state preparation stage, parameter encoding stage and measurement preparation stage, respectively. Subsequently, the output state is measured on a computational basis.}\label{fig:schemes}
\end{figure}

In a typical quantum metrology setup, a probe state $\rho$ is prepared then evolved into $\rho_{\bm{\lambda}}$ through one or more applications of an encoding unitary $U_{\bm{\lambda}}$, which contains $K$ unknown parameters $\bm{\lambda}=(\lambda_1,\lambda_2,\cdots,\lambda_{K})$. Information about $\bm{\lambda}$ can be extracted using a positive operator-valued measurement (POVM) $\{E_x\}$, where $\sum_x E_x=I$. The probability of obtaining a specific measurement outcome $x$ is determined by the Born rule $P(x|\bm{\lambda})={\rm tr}(E_x\rho_{\bm{\lambda}})$. By repeating this measurement process many times, a sequence of outcomes is collected. From this data, an estimate $\bm{\hat{\lambda}}=(\hat{\lambda}_1,\hat{\lambda}_2,\cdots,\hat{\lambda}_{K})$ of the unknown parameters $\bm{\lambda}$ can be derived. Figure~\ref{fig:schemes} illustrates the sequential feedback scheme for this process regarding the multiple uses of $U_{\blamb}$. The green boxes represent quantum gates, while the red circles indicate local noise occurring immediately after each quantum gate due to imperfect quantum operations. We visualize the state and measurement preparation processes with noise in the two gray boxes, respectively, and the output state is subsequently measured on a computational basis. The blue box represents the parameter encoding stage. In this scheme, immediate feedback control, denoted as $V_i$, is allowed after each application of the encoding unitary $U_{\blamb}$.

\begin{figure}[t]
  \centering
  \includegraphics[width=\columnwidth]{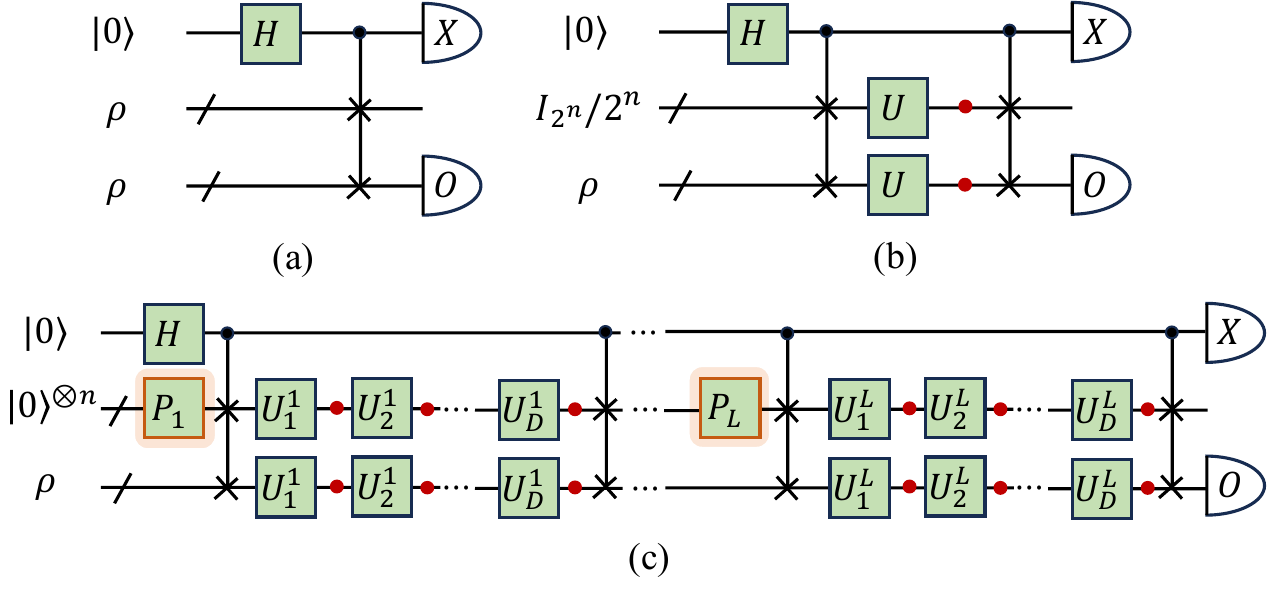}
  \caption{Circuit implementations of virtual purification methods. The 2nd-order implementations for (a) virtual state purification (VSP), (b) virtual channel purification (VCP), and (c) $L$-layer VCP are illustrated. Green boxes denote quantum gates, while red circles indicate noise present in the target quantum circuit. Here, the implementation of VSP and VCP is assumed to be noise-free. Quantum gates $P_i$ in the orange boxes are the tensor product of single-qubit random Pauli unitaries.}\label{fig:circuits_vsp_vcp}
\end{figure}

To suppress the effects of noise in quantum metrology, VSP-based quantum metrology has been proposed to enhance the precision and sensitivity of parameter estimation~\cite{PhysRevLett.129.250503,PhysRevA.109.022410,kwon2025virtualpurificationcomplementsquantum}. Specifically, the $m$th-order VSP utilizes $m$ copies of the target state $\rho$ to measure expectation values with respect to the state
\[\overline{\rho^m}:=\frac{\rho^m}{\tr(\rho^m)}=\frac{\sum_ip_i^m\ket{i}\bra{i}}{\sum_ip_i^m},\]
where $\rho=\sum_ip_i\ket{i}\bra{i}$ represents the spectral decomposition of $\rho$. This approach exponentially suppresses the relative weights of the nondominant eigenvectors in $m$. Figure~\ref{fig:circuits_vsp_vcp}(a) exemplifies the circuit implementation of VSP when $m=2$, where the error-mitigated expectation value of the observable $O$ is given by
\[\frac{\langle X\otimes O\rangle}{\langle X\otimes I_{2^n}\rangle}=\frac{\tr(O\overline{\rho^2})}{\tr(\overline{\rho^2})},\]
with the sampling cost $C_{\rm em}\sim \tr(\overline{\rho^2})^{-2}$~\cite{PhysRevX.11.041036}.
However, it is important to note that VSP can only be applied to the output state. If the circuit is too complex and accumulates significant errors, causing the dominant eigenvector of $\overline{\rho^m}$ to deviate substantially from the noise-free state, VSP-based quantum metrology might not ensure even a constant factor reduction in the bias~\cite{PhysRevA.109.022410}.

To tackle this problem, we now introduce VCP to quantum metrology~\cite{liuVirtualChannelPurification2024}. In practice, suppose a quantum unitary channel $\mathcal{U}$ is followed by the noise channel $\mathcal{E}=p_0\mathcal{I}_{2^n}+\sum_{i=1}^{4^n-1}p_i\overbracket{E_i}$, where $\mathcal{I}_{2^n}$ stands for the $2^n$-dimensional identity channel and $\overbracket{E_i}(\rho)=E_i\rho E_i^{\dagger}$ denotes the channel for a given error component. Crucially, we require $p_0>p_i$ for all $i$, meaning the leading component of $\mathcal{E}$ is the identity channel. Consistent with the assumptions made in VSP~\cite{PhysRevX.11.041036,PhysRevX.11.031057}, the noise channel $\mathcal{E}$ is assumed to be Pauli noise, with $E_i$ being Pauli operations. For general noise, it can be converted into Pauli noise using Pauli twirling~\cite{zcai_paulitwirling19,PhysRevA.94.052325}. Define $\mathcal{U_{E}}=\mathcal{E}\circ\mathcal{U}$. The goal of the $m$th-order VCP is to exploit $m$ copies of $\mathcal{U_{E}}$ to realize $\mathcal{U}_{\mathcal{E}^{(m)}}=\mathcal{E}^{(m)}\circ\mathcal{U}$, where the purified noise channel is of the form
\[\mathcal{E}^{(m)}=\frac{1}{\sum_{i=0}^{4^n-1}p_i^m}\left(p_0^m\mathcal{I}_{2^n}+\sum_{i=1}^{4^n-1}p_i^m\overbracket{E_i}\right).\]
Since the identity channel $\mathcal{I}_{2^n}$ is the dominant component, the noise rate of $\mathcal{E}^{(m)}$ decreases exponentially as $m$ increases. The circuit implementation of VCP with $m=2$ is illustrated in Figure~\ref{fig:circuits_vsp_vcp}(b). Similarly, the error-mitigated expectation value of the observable $O$ is given by
\[\frac{\langle X\otimes O\rangle}{\langle X\otimes I_{2^n}\rangle}=\tr\left(O\mathcal{E}^{(2)}\circ\mathcal{U}(\rho)\right).\]
Let $P_m=\sum_{i=0}^{4^n-1}p_i^m$, the sampling cost is $C_{\rm em}\sim P_m^{-2}$, which is similar to that obtained for VSP. However, compared with VSP, VCP can provide even exponentially stronger error suppression for global noise~\cite{liuVirtualChannelPurification2024}.

For a sequence of quantum operations, instead of applying VCP to the entire circuit, we can adopt a layer-wise implementation of VCP, as illustrated in Figure~\ref{fig:circuits_vsp_vcp}(c). This approach allows for error suppression before it accumulates as the dominant component of the noise channel, thereby circumventing the issues encountered in VSP. Specifically, the control qubit can be reused for each layer. Instead of resetting the ancillary input to the maximally mixed state, random unitary gates can be utilized to achieve the same effect. Notably, since the Pauli group forms a unitary 1-design, we can simply apply random Pauli gates between two layers of VCP to replace the maximally mixed state, as depicted by the orange boxes in Figure~\ref{fig:circuits_vsp_vcp}(c). Moreover, this method of implementing the maximally mixed state does not increase the sample complexity of VCP~\cite{liuVirtualChannelPurification2024}.

\subsection*{Enhanced virtual purification}

Since VCP can mitigate noise before it gets out of control and has a stronger error suppression capability compared to VSP, we introduce it to improve the precision of quantum metrology. However, though VCP is theoretically effective, its practical performance is hindered by noise that occurs during the execution of these protocols. In particular, the controlled-SWAP (cSWAP) gate is notably noisy due to its complex implementation~\cite{PhysRevA.53.2855}. In~\ref{apx:comparison_vcp_vsp}, we examine the performance of these virtual purification methods with and without cSWAP noise in a single-parameter estimation task. The results show that VCP achieves higher precision than VSP, but the benefits of both VCP and VSP are diminished, or can even be negated, in certain scenarios due to this noise. Thus, additional QEM methods should be considered to address the noise in cSWAP gates. The combinations of QEM methods with virtual purification methods 

Several QEM methods have been discussed for incorporation with virtual purification methods~\cite{liuVirtualChannelPurification2024,PhysRevX.11.031057}. To exactly mitigate errors, we adopt PEC to enhance their performance. The sampling cost and the number of different quantum circuits involved in PEC can grow exponentially with the number of noise locations (please refer to~\ref{apx:pec_decomposition} for more details on PEC). In fact, for general noise this exponential growth is theoretically unavoidable for mitigating general noise exactly \cite{PhysRevLett.131.210601,PhysRevLett.131.210602,endo22}. For a $n$-qubit quantum state $\rho$, the $m$th-order VCP introduces noise at ${\mathcal{O}}(mn)$ locations. Although in the sequential scheme of quantum metrology $n={\mathcal{O}}(1)$ usually holds, given limited experimental resources in practice, a natural question arises: Can we reduce the sampling cost and the number of different circuits needed further for the VCP circuit rather than trivially applying PEC to all noise locations? The answer, as it turns out, is yes.

\begin{figure}[!ht]
    \centering
    \includegraphics[width=0.5\textwidth]{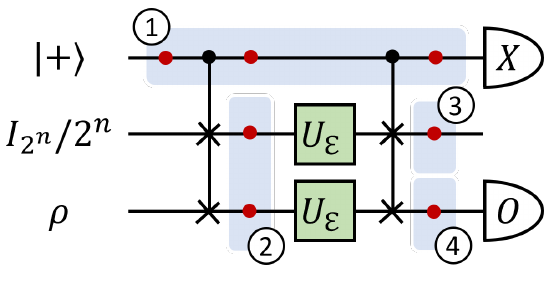}
    \caption{Noise categories for the virtual channel purification circuit. Green boxes represent quantum gates, and red circles indicate local noise. Noise is classified into four types indicated by numbered light blue boxes: \ding{172} noise in the control subsystem, \ding{173} noise in the last two subsystems between two cSWAP layers and \ding{174} (\ding{175}) noise in the ancillary (target) subsystem after the second cSWAP layer.}\label{fig:cswap}
\end{figure}

Take the single-layer VCP circuit with $m=2$ for example, and the insights gained can be naturally extended to larger $m$. The input quantum state in the VCP circuit reads $|+\rangle\langle+|\otimes I_{2^n}/2^n\otimes \rho\in \mathcal{H}_{\rm ctrl}^2\otimes\mathcal{H}_{\rm anc}^{2^n}\otimes\mathcal{H}_{\rm tar}^{2^n}$, where $\mathcal{H}_{\rm ctrl}$, $\mathcal{H}_{\rm anc}^{2^n}$ and $\mathcal{H}_{\rm tar}^{2^n}$ refer to the Hilbert spaces of the control (first), ancillary (second) and target (last) subsystems shown in Figure~\ref{fig:circuits_vsp_vcp} (b), respectively. Assuming local noise models, the noise introduced by the VCP circuit can be divided into four categories: \ding{172} noise in the control subsystem, \ding{173} noise in the last two subsystems between two cSWAP layers and \ding{174} (\ding{175}) noise in the ancillary (target) subsystem after the second cSWAP layer, as illustrated in Figure~\ref{fig:cswap}. For noise in \ding{173}, it is naturally mitigated by the VCP method, and noise in \ding{174} can be ignored since it has no impact on the final result.  Additionally, the noise in \ding{175} only affects the measurement results of the observable $O$. Therefore, its impact on the final result varies depending on the observable $O$, while for an arbitrary observable $O$, additional QEM protocols can be introduced to realize the full benefits of VCP. 

Noise in \ding{172} is the most complex case. Suppose each noise in \ding{172} is characterized by the quantum channel $\mathcal{F}$, and the noise between the cSWAP layers in each of the ancillary and target subsystems is represented by the quantum channel $\mathcal{E}$. Then, according to the following theorem, it turns out that many types of errors in the control subsystem do not introduce systematic error; rather, they only increase statistical error.

\begin{theorem}\label{lemma:ctrl}
    Suppose noise channels $\mathcal{E}=\sum_{i=0}^{4^n-1}p_i\overbracket{E_i}$ and $\mathcal{F}=\sum_{i=0}^3q_i\overbracket{F_i}$ are completely positive trace-preserving (CPTP) channels satisfy the properties
    \begin{equation}\nonumber
    \begin{aligned}
        {\rm tr}(E_iE_j^{\dagger})/2^n=\begin{cases}
            0, & i\neq j\\
            e_i, & i=j
        \end{cases}
    \end{aligned}
    \end{equation}
    and
    \begin{equation}\nonumber
    \begin{aligned}
        \mathcal{F}(|i\rangle\langle j|)=\begin{cases}
            f_{ij}|i\rangle\langle j|, &i\neq j\\
            \sum_k f_{k}^{(i)}|k\rangle\langle k|, & i=j
        \end{cases}
    \end{aligned},
    \end{equation}
    where $e_i\in\mathbb{R}$ and $f_{ij},f_k^{(i)}\in\mathbb{C}$.
    Then, for a local noise channel $\mathcal{F}$ and an $n$-qubit quantum state $\rho$, it holds that
    \begin{equation}\nonumber
        \frac{\langle X\otimes O\rangle_{\tilde{\rho}_{\rm out}}}{\langle X\otimes I_{2^n}\rangle_{\tilde{\rho}_{\rm out}}}=\frac{\langle X\otimes O\rangle_{\rho_{\rm out}}}{\langle X\otimes I_{2^n}\rangle_{\rho_{\rm out}}},
    \end{equation}
    where $\tilde{\rho}_{\rm out}$ and $\rho_{\rm out}$ denote the output states of the virtual channel purification circuit, with and without the existence of $\mathcal{F}$, respectively.
\end{theorem}

By directly calculating the expectation values, it holds that $\langle X\otimes O\rangle_{\tilde{\rho}_{\rm out}}=\eta_m{\rm tr}\left(O\hat{\mathcal{E}}^{(m)}(\rho)\right)
$ and $\langle X\otimes I_{2^n}\rangle_{\tilde{\rho}_{\rm out}}=\eta_m$, where $\hat{\mathcal{E}}^{(m)}=\hat{P}_m^{-1}\left(p_0^me_0^{m-1}\mathcal{I}+\sum_{i=1}^{4^n-1}p_i^me_i^{m-1}\overbracket{E_i}\right)$ with $\hat{P}_m=\sum_{i=0}^{4^n-1}p_i^me_i^{m-1}$, and $\eta_m:={\rm Real}(f_{01}^3)\hat{P}_m$. 
Meanwhile, in the absence of $\mathcal{F}$, the expectation values are similar but with ${\rm Real}(f_{01}^3)=1$. 
Therefore, by dividing these two expectation values, the effect of $\mathcal{F}$ is canceled, resulting in the same result as if $\mathcal{F}$ were not present. Please refer to Supplementary Note~\ref{apx:proof_thrm1} for more details.

Based on the properties discussed in Theorem~\ref{lemma:ctrl}, we can identify specific types of noise channels that satisfy these conditions. For example, $\mathcal{E}$ can be noise channels such as the Pauli channel and amplitude damping channel. Similarly, $\mathcal{F}$ may also encompass noise channels like the amplitude damping channel, as well as the Pauli channel with equal probabilities for $X$ and $Y$ errors, e.g., depolarizing channel and dephasing channel. These noise models are common physical processes that can occur in real quantum systems, hence the analysis can be applied to a wide range of practical scenarios. Therefore, only noise in \ding{175} affects the behaviours of VCP significantly. The numerical simulations in Supplementary Note~\ref{apx:cswap_numerical} verify our analysis.

Furthermore, for multi-layer VCP, the analysis remains applicable for noise in \ding{172}, \ding{173} and \ding{175}. However, noise in \ding{174} cannot be naturally ignored if it occurs in the middle of the circuit. It is important to note that the initial state of the ancillary subsystem for each VCP layer is reset to the maximally mixed state. Specifically, let the quantum state of the ancillary subsystem after noise in \ding{174} be $\rho$. It holds that $\mathbb{E}_P(P\rho P)=I_{2^n}/2^n$, where $P$ is the tensor product of single-qubit random Pauli unitaries~\cite{roy2009unitary}. Therefore, noise in \ding{174} is automatically erased. Additionally, for the noise introduced by performing $P$, when the noise channel $\mathcal{N}$ is unital, such as depolarizing and dephasing channels where $\mathcal{N}(I_d) = I_d$, they have no effect on VCP. Even for nonunital noise channels, such as the amplitude damping channel, since $P$ is the tensor product of single-qubit random Pauli unitaries, with an error rate much lower than that of cSWAP gates, we can simplify the analysis by ignoring this noise.

In summary, for noise channel $\mathcal{F}$ satisfying the condition defined in Theorem~\ref{lemma:ctrl}, only noise in \ding{175} is critical to the performance of VCP. Therefore, PEC can be applied to mitigate noise in \ding{175} for each VCP layer. If the noise channel $\mathcal{F}$ violates the condition, PEC can also be applied to the control subsystem. Given that the control subsystem has a dimension of only 2, and PEC only needs to be applied to the number of locations proportional to the number of VCP layers, the cost of this part can be effectively managed. For simplicity, we primarily focus on the case where the condition holds. As a consequence, the additional cost of applying PEC involves characterizing the noise model, which can be accomplished using quantum process tomography \cite{chuang1997prescription,PhysRevA.87.062119}. Although quantum process tomography generally requires exponential resources with respect to system size, in the context of our task, we only need to focus on the cSWAP gate. This targeted scenario significantly reduces the overhead, making the cost acceptable. In contrast, typical quantum metrology that incorporates prior knowledge of all noise models into the estimator requires quantum process tomography for all quantum operations in the entire quantum metrology process, which may result in exponential cost for characterizing noise models or accumulated error from imperfect noise characterization.

\begin{figure*}[t]
  \centering
  \includegraphics[width=\columnwidth]{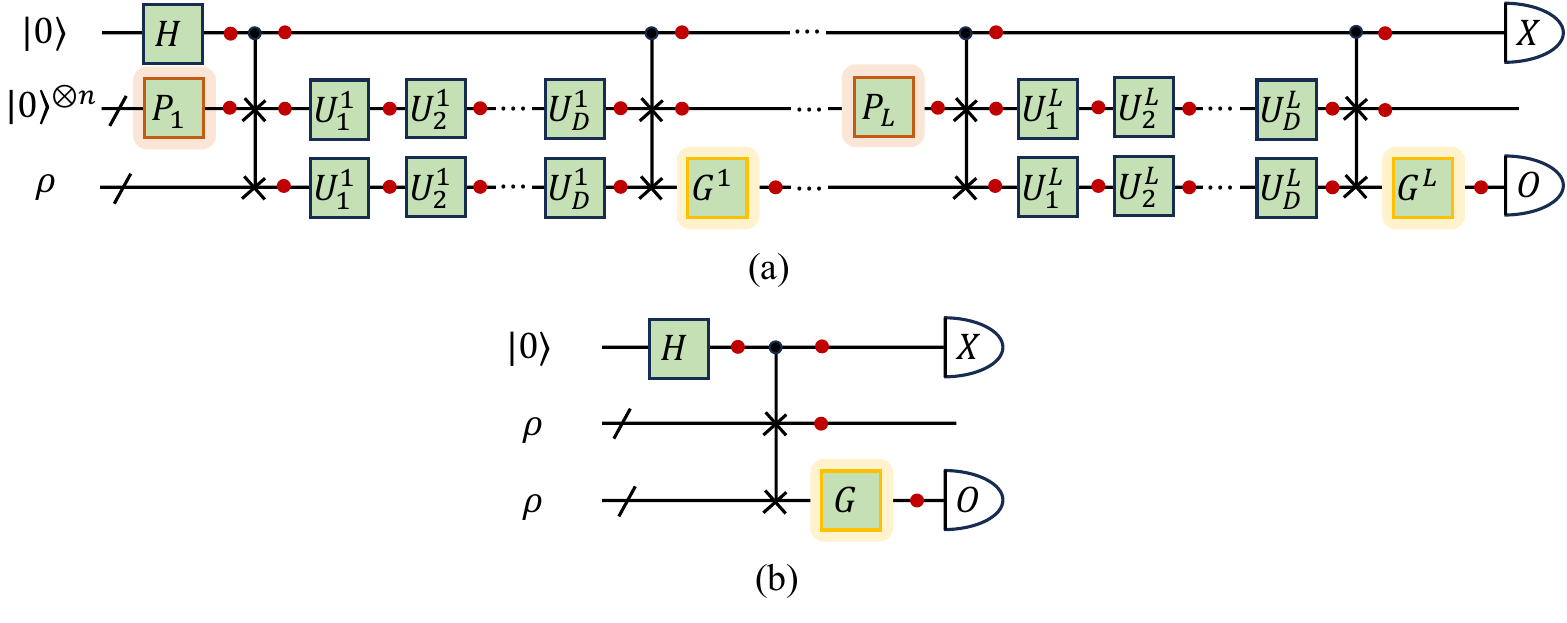}
  \caption{Schemes of enhanced virtual purification that incorporate probabilistic error cancellation (PEC). (a) Enhanced virtual channel purification (VCP-PEC) and (b) enhanced virtual state purification (VSP-PEC). Green boxes denote quantum gates, while red circles indicate noise present in the target quantum circuit. Quantum gates $P_i$ in the orange boxes are the tensor product of single-qubit random Pauli unitaries. Meanwhile, quantum gates $G$ and $G_i$ in the yellow boxes are random gates inserted to cancel the effect of noise in the target subsystem.}\label{fig:vcp_pec}
\end{figure*}

Figure~\ref{fig:vcp_pec}(a) illustrates the enhanced framework for VCP that incorporates PEC, referred to as VCP-PEC. In addition to the layer-wise implementation of VCP shown in Figures~\ref{fig:circuits_vsp_vcp}(c), the quantum gates $G^i$ in the yellow boxes are performed on the target subsystem to mitigate errors via PEC. Please refer to~\ref{apx:pec_decomposition} for more details on the optimal formation of $G^i$ for several common types of noise channels. Additionally, the above conclusion also applies to VSP, as shown in Supplementary Note~\ref{apx:vsp_ctrl}. Therefore, the performance of VSP can also be enhanced using the circuit presented in Figure~\ref{fig:vcp_pec}(b), which is referred to as VSP-PEC.

\subsection*{Error analysis}~\label{sec:error_analysis}

For different quantum metrology tasks, one common goal is to extract the information of unknown parameters $\blamb$ from the expectation value $\tr(O\rho)$, where $\rho$ encodes $\blamb$ and $O$ is an observable. Let $\hat{O}$ be a biased estimator of the noise-free expectation value $\langle O\rangle_{\rho}$, the corresponding mean squared error (MSE) is defined as
\begin{equation}\nonumber
\begin{aligned}
    {\rm MSE(\hat{O})}&=\mathbb{E}\left[\left(\hat{O}-\langle O\rangle_{\rho}\right)^2\right]={\rm Bias}(\hat{O})^2+{\rm Var}(\hat{O})
\end{aligned}
\end{equation}
with ${\rm Bias}(\hat{O})=\mathbb{E}[\hat{O}]-\langle O\rangle_{\rho}$ and ${\rm Var}(\hat{O})=\mathbb{E}[\hat{O}^2]-\mathbb{E}[\hat{O}]^2$. 

Let $\rho=\mathcal{U}_D\circ\cdots\circ\mathcal{U}_1(\rho_{\rm in})$, their noisy implementation is represented by $\tilde{\rho}=\mathcal{E}_D\circ\mathcal{U}_D\circ\cdots\circ\mathcal{E}_1\circ\mathcal{U}_1(\rho_{\rm in})$, where each noise channel $\mathcal{E}_i$ is of an error rate $p_i$. Notice that $\mathcal{U}\circ\mathcal{E}=\mathcal{E}'\circ\mathcal{U}$, where $\mathcal{E}'=\mathcal{U}\circ\mathcal{E}\circ\mathcal{U}^{\dagger}$. Therefore, we can iteratively apply this relation to delay all noise channels at the end, i.e., $\tilde{\rho}=\mathcal{E}_{\rm tot}(\rho)$. For simplicity, we assume the error components of $\mathcal{E}_{\rm tot}$ satisfy the orthogonality defined in Theorem~\ref{lemma:ctrl}, and the noise-free probability is approximated as $p_{\rm ideal}=\prod_i(1-p_i)$.

In single-layer $m$th-order VCP-PEC, for noise channels satisfying Theorem~\ref{lemma:ctrl} the noise increasing the systematic error only exists between the cSWAP layers, which can be merged into the noise channel $\mathcal{E}_{\rm tot}$. Hence, the noise-free probability is increased from $p_{\rm ideal}$ to $p_{\rm ideal}^{{\rm VCP-}m}=p_{\rm ideal}^m \hat{P}_m^{-1}$. 
Then, we have
\begin{equation}\label{eq:main_bias}
\begin{aligned}
    |{\rm Bias}(\hat{O})|&=|\tr\left(O(\tilde{\rho}-\rho)\right)|\\
    &\le \Vert O\Vert_{\infty}\Vert\tilde{\rho}-\rho\Vert_1\\
    &=(1-p_{\rm ideal}^{{\rm VCP-}m})\Vert O\Vert_{\infty}\Vert\hat{\rho}-\rho\Vert_1\\
    &\le 2(1-p_{\rm ideal}^{{\rm VCP-}m})\Vert O\Vert_{\infty}
\end{aligned},
\end{equation}
where $\hat{\rho}$ is defined by $\tilde{\rho}=p_{\rm ideal}^{{\rm VCP-}m}\rho+(1-p_{\rm ideal}^{{\rm VCP-}m})\hat{\rho}$.

Furthermore, recall that VCP-PEC constructs the estimator by division, specifically $\frac{\sum_i\alpha_i\langle X_i\rangle_{\tilde{\rho}_0}}{\sum_i \alpha_i\langle Y_i\rangle_{\tilde{\rho}_0}}$. Here, the observables $X\otimes O$ and $X\otimes I_{2^n}$ are modified to $X_i$ and $Y_i$, respectively. These modified observables are performed on a same quantum state $\tilde{\rho}_0$, ensuring that $\langle X_i\rangle_{\tilde{\rho}_0}=\langle X\otimes O\rangle_{\tilde{\rho}_i}$ and $\langle Y_i\rangle_{\tilde{\rho}_0}=\langle X\otimes I_{2^n}\rangle_{\tilde{\rho}_i}$, where $\tilde{\rho}_i$ denotes the output state of the $i$-th VCP-PEC circuit. The variance of this estimator can be approximated by
\begin{equation}\label{eq:main_var_x/y}
\begin{aligned}
    {\rm Var}\left(\frac{x}{y}\right)\approx\frac{\mu_x^2}{\mu_y^2}\left(\frac{{\rm Var}(x)}{\mu_x^2}-2\frac{{\rm Cov}(x,y)}{\mu_x\mu_y}+\frac{{\rm Var}(y)}{\mu_y^2}\right)
\end{aligned},
\end{equation}
where $x$ and $y$ stands for the estimators of $\sum_i\alpha_i\langle X_i\rangle_{\tilde{\rho}_0}$ and $\sum_i \alpha_i\langle Y_i\rangle_{\tilde{\rho}_0}$, respectively, with expectation values $\mu_x=\eta_m{\rm tr}(O\hat{\mathcal{E}}^{(m)}(\rho))$ and $\mu_y=\eta_m$. Particularly, notice that observables $X\otimes O$ and $X\otimes I_{2^n}$ commute with each other, so they can be measured simultaneously in each circuit run. Hence, we assume that both the nominator and the denominator are estimated using $\nu$ circuit runs. By calculating the corresponding variances and covariance in Eq.~\eqref{eq:main_var_x/y}, it can be derived that the sampling cost required to limit the variation to be $\epsilon^2$ for a bounded observable $O$ is $\nu={\mathcal{O}}\left(\frac{\gamma^2}{\epsilon^2\eta_m^2}\right)$, where $\gamma$ is value related to PEC that grows exponentially with the number $n$ of qubits of $\rho$. Particularly, in the sequential scheme of quantum metrology, $n$ is often kept constant. For more details, please refer to~\ref{apx:variance}.

Additionally, the analysis of variance in the~\ref{apx:variance} quantifies the statistical error introduced by each noise in the control subsystem. As discussed in~\ref{apx:sampling_comparison}, the optimal cost to mitigate the noise can be higher than simply ignoring the noise. This finding is reasonable, as the sampling cost for QEM mentioned earlier is designed to handle arbitrary circuits, whereas the sampling cost we derived applies specifically to virtual purification-based quantum circuits. Nonetheless, this observation further underscores the efficiency of our protocol.

In summary, as $m$ increases, ${\rm Bias}(\hat{O})^2$ approaches zero while the variance increases correspondingly. This establishes a fundamental trade-off between bias reduction and variance control in ${\rm Bias}(\hat{O})^2$ and ${\rm Var}(\hat{O})$. Notice that for general noise, unbiased estimation requires ${\rm Var}(\hat{O})$ to grow exponentially with respect to the number of noisy quantum operations~\cite{PhysRevLett.131.210601,PhysRevLett.131.210602,endo22}, which can easily overwhelm the polynomially enhanced precision in quantum metrology. Therefore, a meaningful question emerges: Can VCP-PEC achieve significant bias reduction while preserving the quantum advantage? Our subsequent results demonstrate that this is indeed possible.

\section*{Results}
\subsection*{Quantum advantage in error-mitigated quantum metrology}
\begin{figure}[t]
  \centering
  \includegraphics[width=\columnwidth]{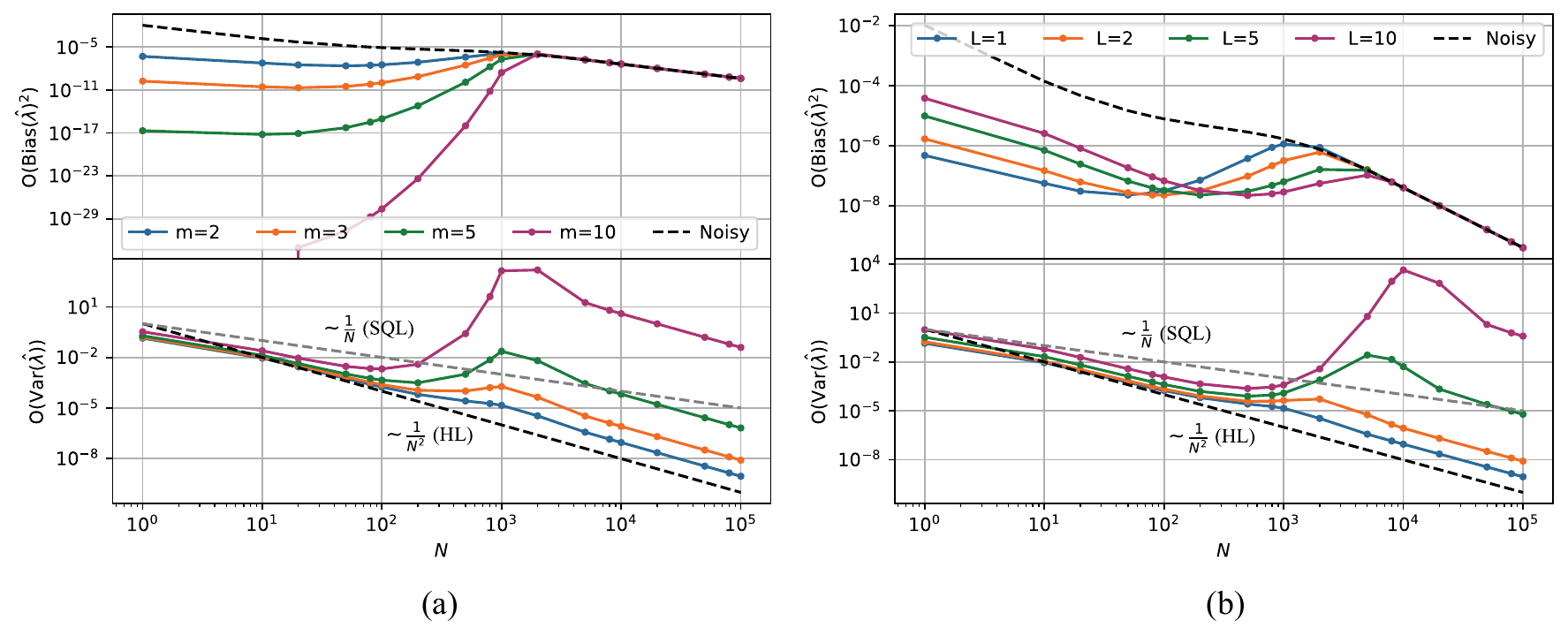}
  \caption{Scaling of the bias and variance of the estimator $\hat{\lambda}$. The results are presented for (a) various orders $m$ with single-layer enhanced virtual channel purification (VCP-PEC) and (b) varying numbers of VCP-PEC layers with $m=2$ across different values of $N$, which is the number of utilized encoding channels. Particularly, the scaling of the standard quantum limit (SQL) and the Heisenberg limit (HL) are depicted by grey and black dashed lines, respectively, in the bottom subfigures of (a) and (b) to benchmark the performance of VCP-PEC.}\label{fig:mse}
\end{figure}

In quantum metrology, the standard quantum limit (SQL) describes the error scaling with the number $N$ of utilized encoding channels that is proportional to $1/N$. A faster decrease in MSE with $N$ presents a quantum advantage. To observe the scaling of ${\rm Bias}(\hat{O})^2$ and ${\rm Var}(\hat{O})$ in $N$, a local single-parameter estimation task is considered.  Assume that the estimator $\hat{\lambda}$ of $\lambda$ is proportional to $\hat{O}/N$~\cite{PhysRevLett.129.250503,PhysRevA.109.022410}. In this case, the values of ${\rm Bias}(\hat{\lambda})^2$ and ${\rm Var}(\hat{\lambda})$ scale as ${\rm Bias}(\hat{O})^2/N^2$ and ${\rm Var}(\hat{O})/N^2$, respectively. Let $n=1$, and suppose that the error rates for single-qubit gates and cSWAP gates are 0.001 and 0.05, respectively. Figure~\ref{fig:mse}(a) illustrates the scaling behavior of ${\rm Bias}(\hat{\lambda})^2$ and ${\rm Var}(\hat{\lambda})$ based on Eqs.~\eqref{eq:main_bias} and~\eqref{eq:main_var_x/y}, with the black dashed line representing the noisy scenario without QEM for comparison. When $N$ is less than $10^3$, VCP-PEC notably reduces the bias. For $m=2,3$, the variance scaling continues to demonstrate a quantum advantage compared to the SQL, indicated by the grey dashed line. However, since $(1-p)^N\approx {1-pN}$ for small $p$, when $N$ exceeds $10^3$, the noise-free probability $p_{\rm ideal}$ can no longer dominate the noise channel $\mathcal{E}_{\rm tot}$. Therefore, the advantage of single-layer VCP-PEC in bias gradually vanishes, and the corresponding variance reaches its maximum value.

To further enhance the performance of VCP-PEC, a multi-layer implementation can be employed. The scaling of bias and variance can be determined by extending the previous analysis. Figure~\ref{fig:mse}(b) presents the results for the 2nd-order VCP-PEC. For small $N$, the single-layer implementation is most effective, as additional layers are impeded by noise introduced by the cSWAP gates. However, when $N$ exceeds 100, the multi-layer VCP-PEC outperforms the single-layer one, enabling higher precision in parameter estimation. For larger $L$, the value of $N$ at which the bias of VCP-PEC converges to the noisy case increases, demonstrating the effectiveness of the multi-layer VCP-PEC. Additionally, although the variance (specifically, the value of $\gamma$) scales exponentially with the number of layers $L$, for probe states of constant size, the increase in variance can still be manageable under practical values of $L$.

Therefore, when $N=\mathcal{O}(p^{-1})$, where $p$ is the error rate of the encoding channel, there exists a region where VCP-PEC achieves substantial bias reduction while effectively controlling variance to preserve quantum advantage. For practical implementations, setting $m=2$ with a constant $L$ could be a suitable choice.
 
\subsection*{Numerical simulation of multi-parameter estimation}

In this section, the performance and robustness of enhanced virtual purification methods are evaluated for multi-parameter estimation tasks under various types of noise. In each scenario, we compare the behavior of five methods: the original noisy method, VSP, VCP, VSP-PEC, and VCP-PEC.~\ref{apx:numerical_single} also presents the comparisons of these methods for single-parameter estimation tasks. Our observations indicate that the enhanced virtual purification methods not only significantly improve the precision and sensitivity of quantum metrology, but also showcase the robustness against practical noises. 

\begin{figure*}[t]
  \centering
  \includegraphics[width=\columnwidth]{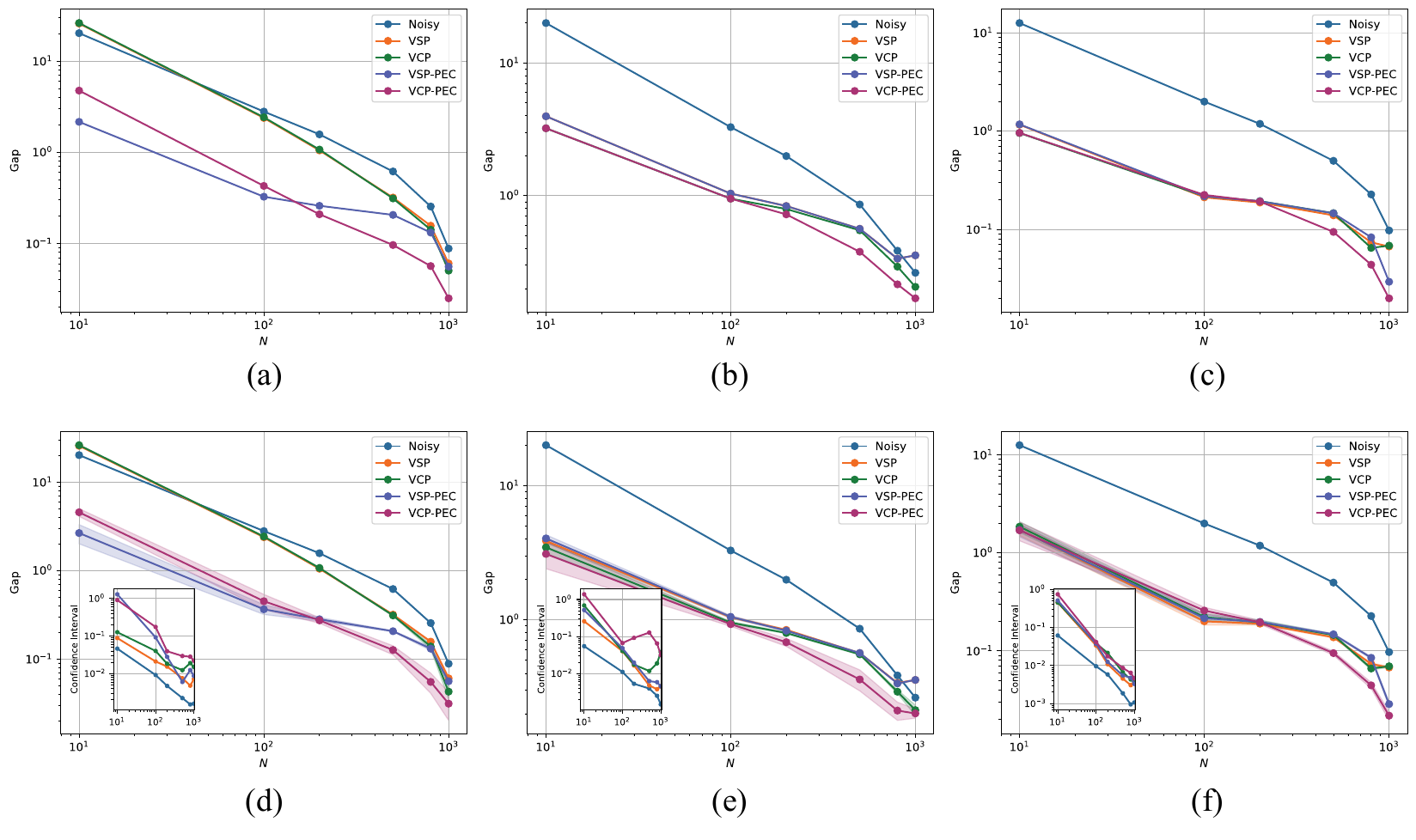}
  \caption{Multi-parameter estimation gaps for different $N$ under different types of noise channels. Here, $N$ represents the number of encoding channels used, and five methods are evaluated: the original noisy method, virtual state purification (VSP), virtual channel purification (VCP), enhanced VSP (VSP-PEC) and enhanced VCP (VCP-PEC).  (a)-(c) plot the performance of different methods with infinite measurement shots, while (d)-(f) depict those with measurement shots $\nu=10^6$. In scenarios with a limited number of measurement shots, experiments are conducted 10 times to calculate the mean values of the gaps and the 95\% confidence intervals. These are represented by solid lines and shaded areas, respectively. Additionally, the exact values of these confidence intervals are provided in the inset subfigures.}\label{fig:exp_multi_seq}
\end{figure*}

Specifically, we consider the Hamiltonian for a spin-1/2 in a magnetic field. The Hamiltonian is written as $H(\bm{\lambda})=B(\sin\theta\cos\phi X+\sin\theta\sin\phi Y+\cos\theta Z)$, where $\bm{\lambda}=(B,\theta,\phi)$ are the unknown parameters to be estimated. Let the probe state be the maximally entangled state $\ket{\psi}=\frac{1}{\sqrt{2}}(\ket{00}+\ket{11})$. The output state evolves under $U_{\bm{\lambda}}=e^{-iH(\blamb)t}\otimes I_2$ for $N$ times and is then measured using a Bell-state measurement. The Bell basis is defined as: $\ket{\phi_1}=\ket{\psi},\ket{\phi_2}=\frac{1}{\sqrt{2}}(\ket{00}-\ket{11}),\ket{\phi_3}=\frac{1}{\sqrt{2}}(\ket{10}+\ket{01})$ and $\ket{\phi_4}=\frac{1}{\sqrt{2}}(\ket{10}-\ket{01})$. In the noise-free case, the probabilities of obtaining each measurement outcome are 
\begin{equation}\label{eq:multi_probs}
\begin{aligned}
    &P(1|\blamb)=\cos^2(BtN),\\
    &P(2|\blamb)=\sin^2(BtN)\cos^2\theta,\\
    &P(3|\blamb)=\sin^2(BtN)\sin^2\theta\cos^2\phi,\\
    &P(4|\blamb)=\sin^2(BtN)\sin^2\theta\sin^2\phi,
\end{aligned}   
\end{equation}
which saturate the quantum Cram\'{e}r-Rao bound~\cite{PhysRevLett.117.160801}.

\subsubsection*{Sequential scheme without feedback}\label{subsec:seq_wo_fb}

Following the experimental setup mentioned above, we first evaluate the performance of different methods using the sequential scheme without feedback. Set $\blamb=(1,0.9,0.8)$ and $t=0.001$. The error rates for single-qubit gates, two-qubit gates, and cSWAP gates are 0.001, 0.01, and 0.05, respectively. By obtaining the measurement outcome probability, the estimate $\hat{\blamb}$ can be derived from Eq.~\eqref{eq:multi_probs}. Define the gap between $\hat{\blamb}$ and $\blamb$ as $\Vert \blamb-\hat{\blamb}\Vert_1$. The performance of VCP and VCP-PEC is presented at the optimal layer $L^*$, with a maximum of 3 layers. 

For $N\in\{10,50,100,200,500, 800,1,000\}$, Figs.~\ref{fig:exp_multi_seq}(a)-(c) depict the performance of these methods under different noise conditions with infinite measurement shots. As observed, the enhanced virtual purification methods significantly outperform other methods in the presence of depolarizing noise. VSP-PEC performs well when $N$ is small; however, as the error accumulates to a substantial level, it is surpassed by VCP-PEC, which aligns with theoretical analysis. For dephasing and amplitude damping noise, VCP-PEC shows superior performance with larger $N$, highlighting its potential for application in complex circuits. 

Furthermore, we also evaluate the behavior of these methods with a limited number of measurement shots, specifically $\nu=10^6$. By repeating the experiments 10 times, Figs.~\ref{fig:exp_multi_seq}(d)-(f) present the mean values of the gaps between the estimated and exact values for each method, depicted by solid lines. The shaded areas represent the corresponding 95\% confidence intervals, with the exact values of these intervals detailed in the inset subfigure. Consistent with theoretical analysis, the error-mitigated estimators are more sensitive to variations in the sampled measurement outcomes, and all four QEM methods exhibit wider confidence intervals than the original noisy quantum circuits. {Specifically, the use of PEC in the enhanced virtual purification methods leads to slightly wider confidence intervals than their original versions. Among them, VCP-PEC shows greater variance than VSP-PEC, as its multi-layer implementation involves multiple applications of PEC.} However, the introduction of PEC does not significantly increase the confidence interval. 

Additionally, in both the infinite and limited measurement shot number cases, the optimal number of layers for VCP is typically 1, whereas for VCP-PEC, the optimal layers are often 2 or 3. By incorporating PEC, the errors introduced by the cSWAP gates are efficiently mitigated, allowing for more VCP layers, which further reduce the error in the target circuit.

\subsubsection*{Sequential feedback scheme}

\begin{figure}[t]
  \centering
  \includegraphics[width=\columnwidth]{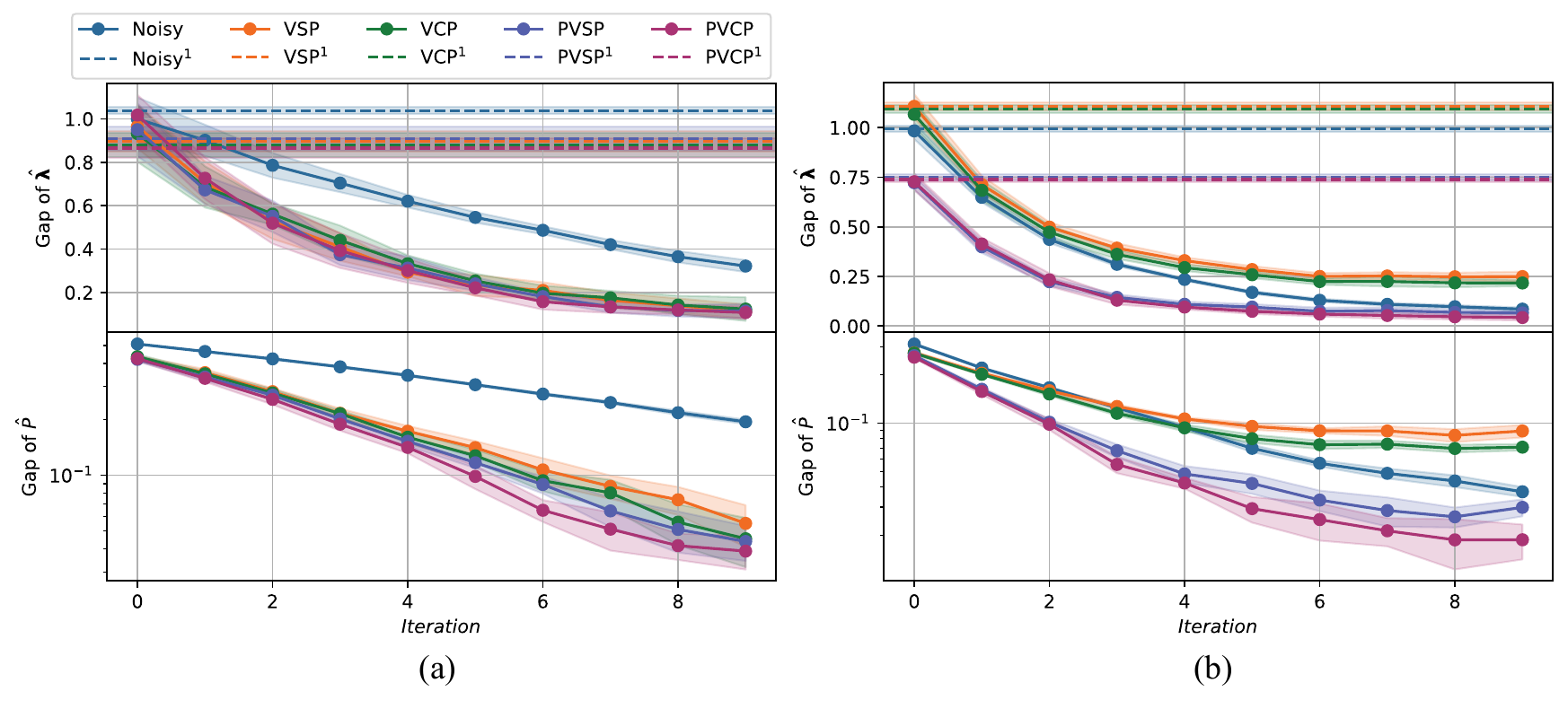}
  \caption{Multi-parameter estimation gaps at different iterations under various types of noise channels. Results are presented for five methods: the original noisy approach, virtual state purification (VSP), virtual channel purification (VCP), enhanced VSP (VSP-PEC), and enhanced VCP (VCP-PEC), under (a) depolarizing noise and (b) amplitude damping noise. For each type of noise channel, the gaps of the estimated measurement outcome probability $\hat{P}_V$ and the estimated parameters $\hat{\blamb}$ derived from $\hat{P}_V$ at each iteration are plotted by solid lines. The number of measurement shots for each method is set to $\nu=10^5$. For comparison, the corresponding estimations obtained from a single iteration optimization with measurement shots $\nu=10^6$ are represented by dashed lines. Each method is repeated 10 times to calculate the mean values of the gaps and the 95\% confidence intervals, which are represented by lines and shaded areas, respectively.}\label{fig:exp_multi_seq_feedback}
\end{figure}

To further enhance the sensitivity of quantum metrology, the sequential feedback scheme is widely adopted. For the multi-parameter estimation task described above, the optimal control $V_i$ after the $i$-th encoding unitary $U_{\blamb}$ has been proved to be $U_{\blamb}^{\dagger}$ for any $i$~\cite{PhysRevLett.117.160801}. Therefore, for simplicity, we denote the optimal control as $V$ without referencing $i$. Since $\blamb$ is not known a priori, $V$ can initially be set to the identity and then updated adaptively as $V=U_{\hat{\blamb}}^{\dagger}$, where $\hat{\blamb}$ is the estimated value of $\blamb$ at each iteration. Instead of obtaining $\hat{\blamb}$ via Eq.~\eqref{eq:multi_probs}, we now estimate $\blamb$ using the maximum likelihood estimator~\cite{380f7170-a649-307c-9495-f3b3298846ff}. Let $\mathbf{x}=(x_1,\cdots,x_{\nu})$ with $x_i\in\{1,2,3,4\}$ represents the sequence of measurement outcomes, and let the model (noise-free) distribution $Q_V(\mathbf{x}|\blamb)$ denote the probability of obtaining measurement outcomes $\mathbf{x}$ given control $V$ and parameters $\blamb$. Since $Q_V(\mathbf{x}|\blamb)$ can be calculated easily, an estimate $\hat{\blamb}$ can be obtained by minimizing the negative log-likelihood loss
\begin{equation}\label{eq:nll}
\begin{aligned}
    &\arg\min_{\hat{\blamb}}\ -\log Q_V(\mathbf{x}|\blamb)\\
    =&\arg\min_{\hat{\blamb}}\ -\nu\sum_{x=1}^{4}\hat{P}_V(x|\blamb)\log Q_V(x|\blamb)
\end{aligned},
\end{equation}
where $\hat{P}_V$ refers to the empirical (noisy) distribution computed from $\mathbf{x}$. By repeatedly replacing $V$ with $U_{\hat{\blamb}}^{\dagger}$ and optimizing Eq.~\eqref{eq:nll} to find a new $\hat{\blamb}$, $\hat{\blamb}$ is expected to converge to $\blamb$.

To observe the behaviors of different methods, we set $\blamb=(\frac{\pi}{4}, \frac{\pi}{6}, \frac{\pi}{6})$, and the output state is measured using a rotated Bell-state measurement by a local operation $e^{i\frac{\pi}{3\sqrt{3}}(X+Y+Z)}$. Additionally, let $N=150,\ t=\frac{1}{2N}$, and suppose the error rates for single-qubit gates, two-qubit gates and cSWAP gates are 0.005, 0.01 and 0.025, respectively. For VCP and VCP-PEC, the performance of the single-layer implementation is valuated.

Set the number of iterations to 10, and each iteration with the measurement shots $\nu=10^5$. By repeating the experiment 10 times, 
Figure~\ref{fig:exp_multi_seq_feedback} depicts the average performance of different methods under different noise channels. Dephasing noise is not considered, as its effect does not vary much for different single-layer virtual purification-based circuits measuring on the computational basis. Specifically, the figure plots the gaps between the estimated measurement outcome probability $\hat{P}_V$ and the noise-free probability $P_V$, i.e., $\Vert\hat{P}_V-P_V\Vert_1$, as well as the gaps between the estimated parameters $\hat{\blamb}$ derived from $\hat{P}_V$ and the true parameters $\blamb$ at each iteration for each type of noise channel, shown as solid lines. The results demonstrate that error-mitigated estimations significantly outperform the results of the original noisy quantum circuits for depolarizing noise.  However, VSP and VCP do not achieve improved performance with QEM under amplitude damping noise, whereas both enhanced virtual purification methods maintain their advantages. In particular, in these tasks VCP-PEC achieves the highest precision for $P_V$, leading to the best estimation of $\blamb$. 

As a comparison, the corresponding estimations from these methods obtained through a single iteration optimization with the same measurement cost, i.e., $\nu=10^6$, are shown by dashed lines. The behavior of these methods generally remains consistent with previous observations, while highlighting the advantage of sequential feedback schemes over those without feedback control.

\subsection*{Robustness}

In the previous analysis, the noise model was assumed to be local; however, in practice, the application of cSWAP gates can involve correlated noise. Furthermore, the implementation of enhanced virtual purification methods requires the characterization of cSWAP gate noise. Due to the imperfections in quantum operations, accurately characterizing the noise model may be infeasible. To evaluate the robustness of our enhanced virtual purification methods against practical noise, two types of noise models are considered for cSWAP gates: $\mathcal{E}_{\rm DP}(p_{0}, p_1)=\circ_{i=1}^{3}\mathcal{E}^{(i)}_{{\rm DP}}(p_0)\circ\mathcal{E}^{\rm global}_{{\rm DP}}(p_1)$ and $\mathcal{E}_{\rm FP}(p_0,p_1)=\circ_{i=1}^{3}\mathcal{E}^{(i)}_{{\rm PF}}(p_0)\circ\mathcal{E}^{\rm global}_{{\rm PF}}(p_1)$. Here, $\mathcal{E}_{\rm DP}^{(i)}(p_0)$ ($\mathcal{E}_{\rm PF}^{(i)}(p_0)$) refers to the local depolarizing (dephasing) channel with error rate $p_0$ acting on $i$-th qubit, while $\mathcal{E}^{\rm global}_{{\rm DP}}(p_1)$ ($\mathcal{E}^{\rm global}_{{\rm PF}}(p_1)$) represents the global depolarizing (global dephasing, i.e., $\mathcal{E}^{\rm global}_{{\rm PF}}(p_1)=(1-p)\mathcal{I}_{2^3}+p\mathcal{Z}^{\otimes 3}$) channel with error rate $p_1$. Particularly, the performances of VSP and VCP are excluded from consideration, as they do not require prior knowledge of the noise model in cSWAP gates.

\begin{figure}[!ht]
  \centering
  \includegraphics[width=0.55\textwidth]{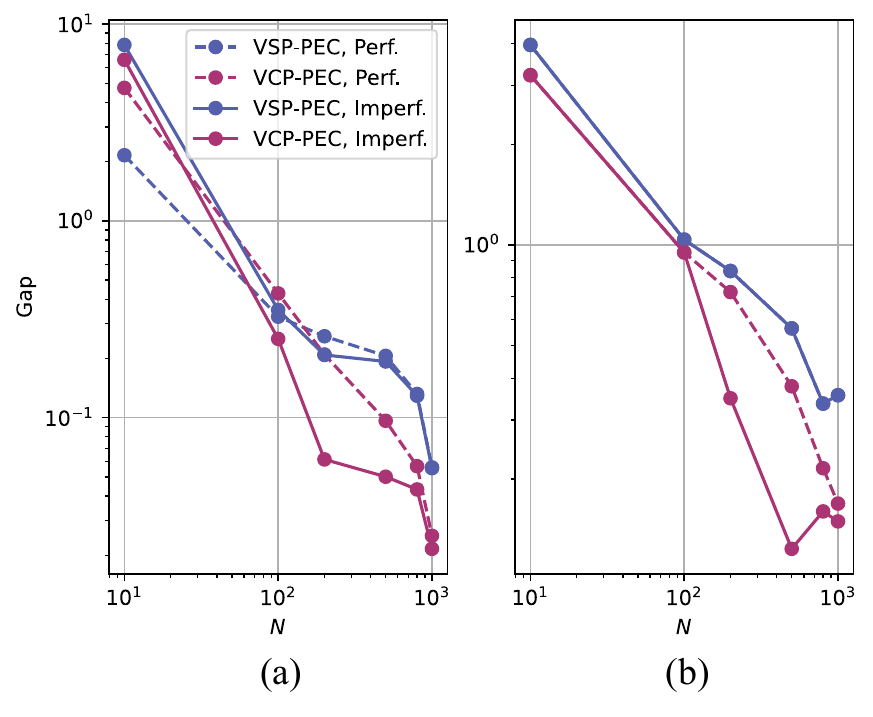}
  \caption{Robustness of enhanced virtual purification methods against cSWAP noise. The parameter estimation gaps for varying numbers $N$ of utilized encoding channels are plotted for enhanced virtual state purification (VSP-PEC) and enhanced virtual channel purification (VCP-PEC) under noise channels (a) $\mathcal{E}_{\rm DP}(0.05, 0.01)$ and (b) $\mathcal{E}_{\rm PF}(0.05, 0.01)$. The dashed lines represent the results when probabilistic error cancellation (PEC) perfectly cancels cSWAP noise, and the solid lines depict the performance with imperfect PEC cancellation. Specifically, for the latter case, PEC realizing the inverse of the corresponding local noise channel with an error rate of 0.055 is applied to cancel the actual local noise with an error rate of 0.05.}\label{fig:robustness}
\end{figure}

For each noise type, we assume a 10\% relative error in estimating the error rate of cSWAP gates. Specifically, let the noise in cSWAP gates be $\mathcal{E}_{\rm DP}(0.05, 0.01)$, and the PEC targeting canceling only the local noise $\circ_{i=1}^{3}\mathcal{E}_{{\rm DP}}^{(i)}(0.055)$ is applied. Analogously for $\mathcal{E}_{\rm PF}(0.05, 0.01)$. Based on the experimental setup described in the ``Sequential scheme without feedback" subsection in the Results, Figure~\ref{fig:robustness} presents the corresponding results. The dashed lines display the results when PEC perfectly cancels cSWAP noise in the target subsystem, as shown in Figures~\ref{fig:exp_multi_seq}(a) and (b), while the solid lines represent the performance of these enhanced virtual purification methods under practical noise with imperfect PEC cancellation of local noise. For VSP-PEC, the gaps change little for depolarizing noise and remain unchanged for dephasing noise, as measurements are taken in the computational basis. Besides, VCP-PEC achieves descent or even more precise parameter estimation for both types of noise. Thus, our enhanced virtual purification methods demonstrate robustness against practical noise, indicating their potential for practical applications.

\section*{Conclusions}

Targeting significant noise accumulation and noisy implementations of QEM protocols,  
VCP-PEC is proposed to achieve more precise estimations for unknown parameters on {near-term} devices. Specifically, VCP-PEC addresses the limitation of VSP in handling substantial noise accumulation, and fully exploits the efficacy of VCP. Moreover, it mitigates errors with a reasonable sampling cost, as the number of noise locations where PEC is applied is well managed. Additionally, our strategy for applying PEC can be naturally adapted for VSP. For shallow circuits, VSP-PEC can achieve comparable performance to VCP-PEC, making it a good choice for practical applications due to its simpler implementation. The efficacy of these enhanced virtual purification methods is systematically evaluated for both single- and multi-parameter estimation tasks under various types of noise channels, where a significant improvement is achieved compared with the original noisy and virtual purification methods. Notably, they also demonstrate robustness against correlated noise and the inexact characterization of noise models, underscoring their significance in practical applications.

Additionally, the scaling of bias and variance for VCP-PEC has been analyzed. It was found that VCP-PEC significantly reduces bias while maintaining a quantum advantage when the number of involved encoding unitaries is less than the inverse of the error rate of the encoding channel. This is consistent with numerical simulations for specific parameter estimation tasks. However, to further suppress noise beyond this range, the quantum advantage in variance scaling can disappear. Thus, incorporating a QEM method with QEC is necessary to achieve a practical quantum advantage~\cite{kwon2025virtualpurificationcomplementsquantum}. We leave this integration for future work. Lastly, it is particularly noteworthy that both enhanced virtual purification methods can be naturally extended beyond quantum metrology, and can therefore be applied to various tasks in quantum information and quantum computation.

\backmatter

\bmhead{Acknowledgements}

We thank Ruiqi Zhang for helpful discussions. This work is supported by the Direct Grant of The Chinese University of Hong Kong (Grant No. 4055260), the Innovation Program for Quantum Science and Technology (2023ZD0300600), the Guangdong Provincial Quantum Science Strategic Initiative (GDZX2303007), the Research Grants Council of Hong Kong (14309223, 14309624,14309022), 1+1+1 CUHK-CUHK(SZ)-GDST Joint Collaboration Fund (Grant No. GRDP2025-022).


\appendix

\renewcommand{\thesection}{Supplementary Note \Alph{section}}
\renewcommand{\thesubsection}{\Alph{section}.\arabic{subsection}}

\renewcommand{\figurename}{Supplementary Fig.}
\renewcommand{\thefigure}{\arabic{figure}} 
\setcounter{figure}{0} 

\renewcommand{\tablename}{Supplementary Table}
\renewcommand{\thetable}{\arabic{table}}
\setcounter{table}{0} 

\setcounter{equation}{0} 
\renewcommand{\theequation}{S\arabic{equation}} 

\section{Comparisons between virtual purification methods}\label{apx:comparison_vcp_vsp}

To demonstrate and compare the efficacy of virtual state purification (VSP) and virtual channel purification (VCP) in mitigating systematic errors, we consider both the sequential scheme mentioned in the main text and the parallel scheme illustrated in Supplementary Figure~\ref{fig:parallel_schemes}. Specifically, a single-parameter estimation task is examined in the presence of three common types of local noise: depolarizing noise, dephasing noise, and amplitude damping noise. 

\begin{figure}[!ht]
  \captionsetup[subfigure]{justification=centering, labelfont=bf}
  \centering
  \begin{subfigure}[!ht]{0.45\textwidth}
    \centering
    \includegraphics[width=\textwidth]{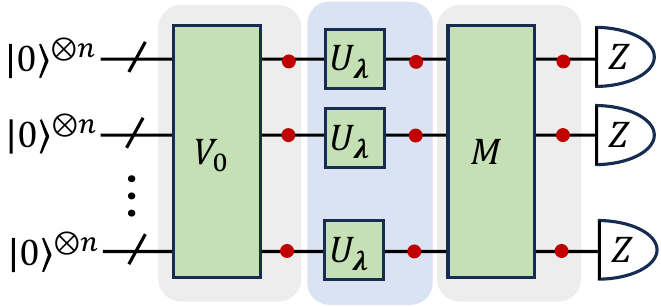}
  \end{subfigure}
  \caption{Parallel scheme of quantum metrology.  Green boxes represent quantum gates, while red circles indicate local noise occurring immediately after each quantum gate. Particularly, the first gray box, the blue box and the final gray box represent the state preparation stage, parameter encoding stage and measurement preparation stage, respectively. Subsequently, the output state is measured on a computational basis.}\label{fig:parallel_schemes}
\end{figure}

Specifically, we consider a uniform magnetic field described by the Zeeman Hamiltonian $H=\sum_{j=1}^n\lambda Z^{(j)}/2$ with unknown parameter $\lambda$ determined by the target field, where $Z^{(j)}$ denotes the Pauli $Z$ operator acting on the $j$-the qubit. In the parallel scheme, the initial probe state is selected as the $n$-qubit Greenberger-Horne-Zeilinger (GHZ) state, represented as $\rho_0=|{\rm GHZ}_n\rangle\langle {\rm GHZ}_n|$, where $|{\rm GHZ}_n\rangle=\frac{1}{\sqrt{2}}(|0\rangle^{\otimes n}+|1\rangle^{\otimes n})$. By measuring $P_y=|{\rm GHZ}_y\rangle\langle{\rm GHZ}_y|$, with $|{\rm GHZ}_y\rangle=\frac{1}{\sqrt{2}}\left(|0\rangle^{\otimes n}-i|1\rangle^{\otimes n}\right)$, we derive $\lambda=\arcsin(1-2\langle P_y\rangle)/n$~\cite{PhysRevLett.129.250503}. Here, we set $n=N$. Similarly, in the sequential scheme, we use the same setting as the parallel scheme with $n=1$, except that the Hamiltonian $H$ is applied repeatedly $N$ times.

\begin{figure*}[t]
  \captionsetup[subfigure]{justification=centering, labelfont=bf}
  \centering
  \begin{subfigure}[t]{0.33\textwidth}
    \centering
    \includegraphics[width = \linewidth]{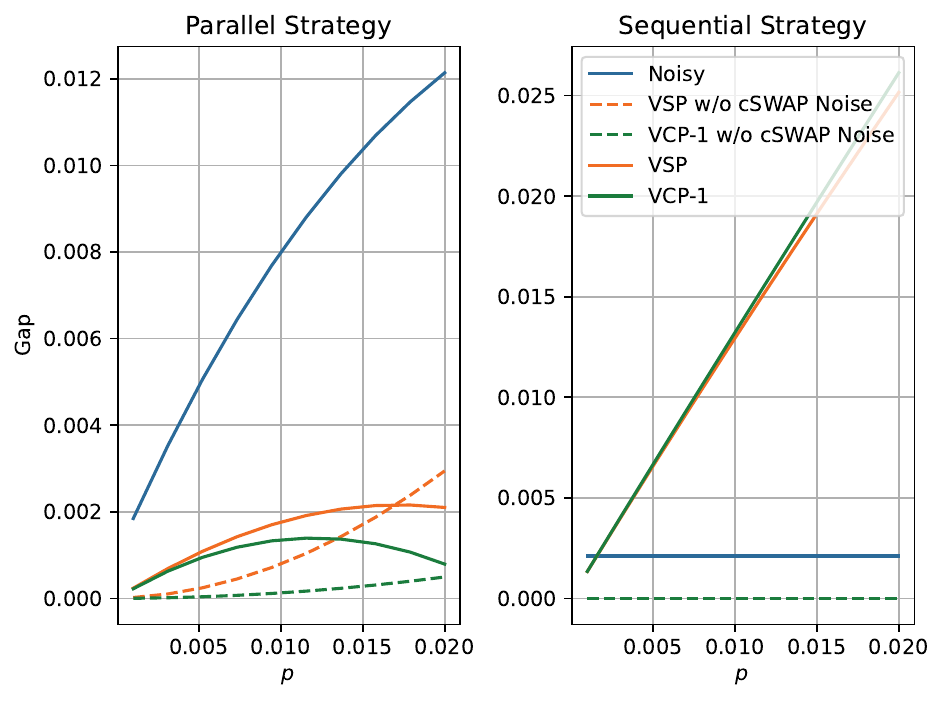}
    \subcaption{Depolarizing noise}
  \end{subfigure}%
  \hfill
  \begin{subfigure}[t]{0.33\textwidth}
    \centering
    \includegraphics[width = \linewidth]{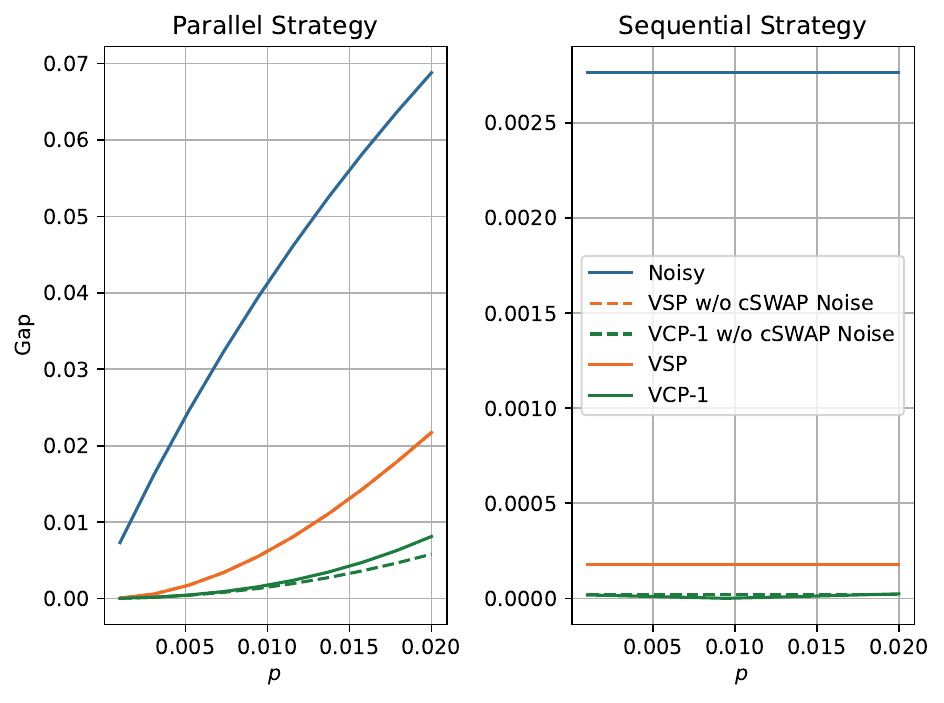}
    \subcaption{Dephasing noise}
  \end{subfigure}%
  \hfill
  \begin{subfigure}[t]{0.33\textwidth}
    \centering
    \includegraphics[width = \linewidth]{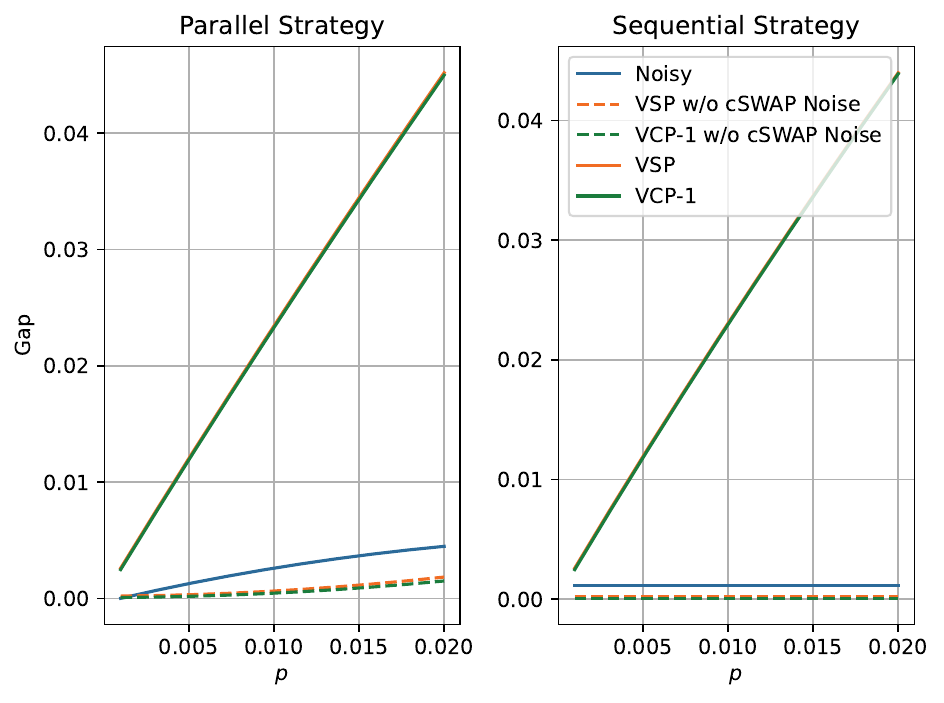}
    \subcaption{Amplitude damping noise}
  \end{subfigure}%

  \caption{Parameter estimation gaps under varying noise levels $p$ for local (a) depolarizing noise, (b) dephasing noise, and (c) amplitude damping noise. The error rates for single-qubit and two-qubit gates are 0.001 and $p\in[0.001,0.02]$. The gaps in the original quantum circuit, as well as those after applying VSP and VCP-1 under different $p$, are represented by the blue solid line, orange and green dashed lines, respectively. Besides, when noisy cSWAP gates are introduced, with an error rate of $5p$, the gaps following VSP and VCP-1 for different $p$ are depicted by orange and green solid lines, respectively.}\label{fig:vcp_vs_vsp}
\end{figure*}

With $N=5$ and $m=2$, VSP and single-layer VCP, referred to as VCP-1, are implemented using the quantum circuits depicted in Figure 2(a) and (b) in the main text, respectively. In these experiments, the error rates for single-qubit and two-qubit gates are set at 0.001 and $p\in[0.001,0.02]$, respectively. Supplementary Figure~\ref{fig:vcp_vs_vsp} presents the resulting estimation errors, denoted as the gaps $|\lambda-\hat{\lambda}|$, obtained by applying VSP and VCP-1 without noise in the controlled-SWAP (cSWAP) gates (illustrated by the orange and green dashed lines, respectively). As the results indicate, both VSP and VCP significantly reduce errors compared to those obtained from the original noisy setup (represented by the blue solid line). Furthermore, VCP consistently demonstrates superior performance over VSP.

However, in practice, the cSWAP gate is notably noisy because its implementation requires at least five two-qubit gates~\cite{PhysRevA.53.2855}. Therefore, it is essential to evaluate the performance of these virtual purification methods under conditions where cSWAP gates are subject to noise. Specifically, the error rate for the cSWAP gates is set to $5p$, five times greater than that of two-qubit gates. As illustrated in Supplementary Figure~\ref{fig:vcp_vs_vsp}, when noise is introduced into the cSWAP gates, the benefits of using VSP and VCP (depicted by the orange and green solid lines, respectively) are diminished or can even be negated in certain tasks. Thus, additional operations should be implemented to mitigate errors in cSWAP, thereby reducing errors in practical applications.

\section{Probabilistic error cancellation}\label{apx:pec_decomposition}

Probabilistic error cancellation (PEC) is a typical QEM protocol that inverts well-characterized noise channels, thereby canceling the effect of noise to obtain an unbiased estimation of the noise-free expectation~\cite{PhysRevLett.119.180509,PhysRevX.8.031027}. Specifically, for an ideal quantum unitary channel $\mathcal{U}$, let $\{\mathcal{B}_i\}$ be a sufficiently large set of noisy quantum operations that can span it. This means we can decompose $\mathcal{U}$ as $\mathcal{U}=\sum_i\alpha_i\mathcal{B}_i$, where $\alpha_i$ are real numbers satisfying $\sum_i\alpha_i=1$. Then, for a target observable $O$ on the quantum state $\mathcal{U}(\rho)$, it holds that
\begin{equation}\label{eq:pec}
    \langle O\rangle=\sum_i\alpha_i\tr\left(O\mathcal{B}_i(\rho)\right)=\sum_i p_i\cdot\gamma{\rm {sign}}(\alpha_i)\cdot \tr\left(O\mathcal{B}_i(\rho)\right),
\end{equation}
where $\gamma=\sum_i|\alpha_i|\ge 1$, $p_i=\frac{|\alpha_i|}{\gamma}$, and ${\rm sign(\cdot)}$ indicates the sign of the real number. According to this equation, we can execute $\mathcal{B}_i$ with probability $p_i$, and then multiply the corresponding measurement outcome by $\gamma{\rm sign}(\alpha_i)$ to obtain an estimate. By repeating this procedure many times, the average value converges to the expectation value $\langle O\rangle$. To limit the variance to $\epsilon^2$, the number of trials required is ${\mathcal{O}}(\gamma^2/\epsilon^2)$. More generally, to construct a sequence of ideal unitary channels $\mathcal{U}=\circ_{k=0}^{K-1} \mathcal{U}_k$, suppose each unitary channel can be decomposed into $\mathcal{U}_k=\sum_i\alpha_{k,i}\mathcal{B}_{k,i}$. Then, we have
\begin{equation}\nonumber
    \langle O\rangle=\sum_{\bm{i}}\alpha_{\bm{i}}\tr\left(O\mathcal{B}_{\bm{i}}(\rho)\right)=\sum_{\bm{i}} p_{\bm{i}}\cdot\gamma{\rm {sign}}(\alpha_{\bm{i}})\cdot \tr\left(O\mathcal{B}_{\bm{i}}(\rho)\right),
\end{equation}
where $\bm{i}=(i_0,i_1,\cdots,i_{K-1})$, $\alpha_{\bm{i}}=\prod_{k=0}^{K-1} \alpha_{k,i_k}$, $\mathcal{B}_{\bm{i}}=\circ_{k=0}^{K-1}\mathcal{B}_{k,i_k}$, and $\gamma=\sum_{\bm{i}}|\alpha_{\bm{i}}|=\prod_k\gamma_k$ with $\gamma_k=\sum_{i_k}|\alpha_{k,i_k}|$. Hence, the sampling cost grows exponentially with $K$. The estimation of the expectation value follows the same procedure as in the single noise case. 

\begin{table*}\footnotesize
    \centering
    \caption{Optimal decompositions for different noise channels $\mathcal{E}$ with noise level $p$~\cite{PhysRevResearch.3.033178}. For an ideal quantum unitary channel $\mathcal{U}$, it can be decomposed into $\mathcal{U}=\sum_i\alpha_i\mathcal{B}_i$. Quantum channels $\mathcal{X},\mathcal{Y},\mathcal{Z}$ stand for Pauli operators, and $\mathcal{P}_{\ket{0}}$ denotes the zero-state preparation channel.}
    \begin{tabular}{c|cc}
    \hline
    \hline
     & $\bm{\alpha}$ & $\{\mathcal{B}_i\}$\\
    \hline
    Depolarizing & $\left(1+\frac{3p}{4(1-p)}, -\frac{p}{4(1-p)}, -\frac{p}{4(1-p)}, -\frac{p}{4(1-p)}\right)$ & $\{\mathcal{E}\circ\mathcal{U},\mathcal{E}\circ\mathcal{X}\circ\mathcal{U},\mathcal{E}\circ\mathcal{Y}\circ\mathcal{U},\mathcal{E}\circ\mathcal{Z}\circ\mathcal{U}\}$\\
    Dephasing & $\left(\frac{1-p}{1-2p}, -\frac{p}{1-2p}\right)$ & $\{\mathcal{E}\circ\mathcal{U},\mathcal{E}\circ\mathcal{Z}\circ\mathcal{U}\}$\\
    Amplitude damping & $\left(\frac{1+\sqrt{1-p}}{2(1-p)},\frac{1-\sqrt{1-p}}{2(1-p)},-\frac{p}{1-p}\right)$ & $\{\mathcal{E}\circ\mathcal{U},\mathcal{E}\circ\mathcal{Z}\circ\mathcal{U},\mathcal{E}\circ\mathcal{P}_{\ket{0}}\}$\\
    \hline
    \hline
    \end{tabular}
    \label{tab:pec}
\end{table*}

According to the assumptions made in PEC~\cite{PhysRevResearch.3.033178}, operators in the noisy basis $\{\mathcal{B}_i\}$ can be of the form $\mathcal{E}\circ \mathcal{G}\circ\mathcal{U}$, where $\mathcal{G}$ is a quantum operation. 
For example, assume the local noise is the depolarizing channel $\mathcal{E}_{DP}$ with error rate $p$. The optimal decomposition of the ideal unitary channel $\mathcal{U}$ with respect to the minimal sampling cost is $\mathcal{U}=\sum_i\alpha_i\mathcal{E}_{DP}\circ\mathcal{G}_i\circ\mathcal{U}$, where $\alpha_i$ equal to $1+\frac{3p}{4(1-p)}$ for $\mathcal{G}_0=\mathcal{I}_2$ and equal to $-\frac{p}{4(1-p)}$ for $\mathcal{G}_{1,2,3}=\mathcal{X},\mathcal{Y},\mathcal{Z}$. 
Supplementary Table~\ref{tab:pec} lists the corresponding coefficients $\alpha_i$ and noisy quantum operations $\mathcal{B}_i$ that form the optimal decomposition of three common types of noise channels with respect to the minimal sampling cost.

\section{Impact of noise in cSWAP gates}
\subsection{Proof of Theorem~\ref{lemma:ctrl}}\label{apx:proof_thrm1}

First, we consider the 2nd-order VCP circuit. Suppose the channel of noisy operation is $\mathcal{U}_{\mathcal{E}}=\mathcal{E}\circ\mathcal{U}$, we then have
\begin{equation}\nonumber
\begin{aligned}
    &\overbracket{\rm cSWAP}\circ\left(\mathcal{I}_2\otimes\mathcal{U}^{\otimes2}_{\mathcal{E}}\right)\circ\overbracket{\rm cSWAP}\\
    =&\overbracket{\rm cSWAP}\circ\left(\mathcal{I}_2\otimes\mathcal{E}^{\otimes2}\right)\circ\overbracket{\rm cSWAP}\circ\left(\mathcal{I}_2\otimes\mathcal{U}^{\otimes2}\right)
\end{aligned},
\end{equation}
where $\overbracket{\rm cSWAP}(\cdot)={\rm cSWAP}(\cdot){\rm cSWAP}^{\dagger}$ denotes the cSWAP channel. Therefore, we can examine the impact of noise in the control subsystem using the circuit depicted in Supplementary Figure~\ref{fig:ctrl} for simplicity.

\begin{figure}[!ht]
    \centering
    \includegraphics[width=0.45\textwidth]{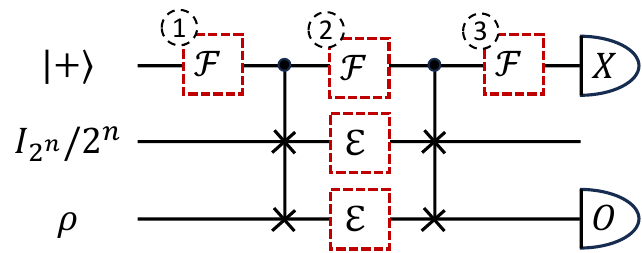}
    \caption{Three locations where noise channel $\mathcal{F}$ can occur in the control subsystem.}\label{fig:ctrl}
\end{figure}

First, consider $\mathcal{F}$ occurring only in the first two locations in Supplementary Figure~\ref{fig:ctrl}. The state immediately before the second cSWAP gate is given by
\begin{equation}\label{eq:ctrl12}
\begin{aligned}
    \tilde{\rho}&=(\mathcal{F}\otimes\mathcal{E}\otimes\mathcal{E})\circ\overbracket{\rm cSWAP}\circ\left(\sigma\otimes I_{2^n}/2^{n}\otimes\rho\right)\\
    &\begin{aligned}
        =2^{-n}\left(\right.&\sigma_{00}\mathcal{F}(|0\rangle\langle0|)\otimes\mathcal{E}^{\otimes 2}(I_{2^n}\otimes\rho)\\
        &+\sigma_{01}\mathcal{F}(|0\rangle\langle1|)\otimes\mathcal{E}^{\otimes 2}((I_{2^n}\otimes\rho){\rm SWAP})\\
        &+\sigma_{10}\mathcal{F}(|1\rangle\langle0|)\otimes \mathcal{E}^{\otimes 2}({\rm SWAP}(I_{2^n}\otimes\rho))\\
        &+\sigma_{11}\mathcal{F}(|1\rangle\langle1|)\otimes\mathcal{E}^{\otimes 2}(\rho\otimes I_{2^n})\left.\right)
    \end{aligned}
\end{aligned},
\end{equation}
where $\sigma:=\mathcal{F}(|+\rangle\langle+|)$ and $\sigma_{ij}$ is the $(i,j)$-th entry of $\sigma$. The terms $\mathcal{F}(|0\rangle\langle0|)$ and $\mathcal{F}(|1\rangle\langle1|)$ can only expand $\{|0\rangle\langle0|,|1\rangle\langle1|\}$. Since the cSWAP gate does not change their basis, their expectation values $\langle X\rangle$ of the control qubit measurement equal zero. Then, we can only focus on the middle two terms in Eq.~\eqref{eq:ctrl12}. 

Define 
\[\hat{\sigma}:=\overbracket{\rm cSWAP}\left(\mathcal{F}(|0\rangle\langle1|)\otimes\mathcal{E}^{\otimes 2}((I_{2^n}/2^n\otimes\rho){\rm SWAP})\right),\] 
the corresponding measurement result is
\begin{equation}\nonumber
\begin{aligned}
    \langle X\otimes O\rangle_{\hat{\sigma}}&=
    \sum_{i,j=0}^{4^n-1}\frac{p_ip_j}{2^n}
    \left[f_{01}\langle{1|X|0\rangle}{\rm tr}(OE_i\rho E_j^{\dagger}){\rm tr}(E_j I_{2^n} E_i^{\dagger})\right]\\
    &=f_{01}\sum_{i=0}^{4^n-1}p_i^2e_i{\rm tr}(OE_i\rho E_i^{\dagger})\\
    &=f_{01}\cdot \hat{P}_2{\rm tr}\left(O\hat{\mathcal{E}}^{(2)}\rho\right)
\end{aligned},
\end{equation}
where $\hat{\mathcal{E}}^{(m)}=\hat{P}_m^{-1}\left(p_0^me_0^{m-1}\mathcal{I}+\sum_{i=1}^{4^n-1}p_i^me_i^{m-1}\overbracket{E_i}\right)$ with $\hat{P}_m=\sum_{i=0}^{4^n-1}p_i^me_i^{m-1}$. Here, the second equality adopts the facts that ${\rm tr}(E_iE_j^{\dagger})=0$ for all $i\neq j$ and ${\rm tr}(E_iE_i^{\dagger})/2^{n}=e_i$. Similarly, for 
\[\hat{\sigma}':=\overbracket{\rm cSWAP}\left(\mathcal{F}(|1\rangle\langle0|)\otimes\mathcal{E}^{\otimes 2}({\rm SWAP}(I_{2^n}/2^n\otimes\rho))\right),\]
we have
\begin{equation}\nonumber
\begin{aligned}
    \langle X\otimes O\rangle_{\hat{\sigma}'}&=f_{10}\cdot \hat{P}_2{\rm tr}\left(O\hat{\mathcal{E}}^{(2)}(\rho)\right)
\end{aligned}.
\end{equation}
Consequently, for the state $\overbracket{\rm cSWAP}(\tilde{\rho})$, we have
\begin{equation}\nonumber
\begin{aligned}
    \langle X\otimes O\rangle_{\overbracket{\rm cSWAP}(\tilde{\rho})}
    &=\sigma_{01}\langle X\otimes O\rangle_{\hat{\sigma}}+\sigma_{10}\langle X\otimes O\rangle_{\hat{\sigma}'}\\
    &=(\sigma_{01}f_{01}+\sigma_{10}f_{10})\cdot \hat{P}_2{\rm tr}\left(O\hat{\mathcal{E}}^{(2)}(\rho)\right)\\
    &={\rm Real}(f_{01}^2)\hat{P}_2{\rm tr}\left(O\hat{\mathcal{E}}^{(2)}(\rho)\right)
\end{aligned},
\end{equation}
where ${\rm Real}(\cdot)$ stands for the real part of the complex number. The last equality holds since $\sigma_{01}=\frac{1}{2}f_{01}=\sigma_{10}^*$, where $*$ denotes the conjugate. 

Then, suppose $\mathcal{F}$ also occurs in the last location, it can be easily checked that
\begin{equation}\label{eq:ctrl_xo}
\begin{aligned}
    \langle X\otimes O\rangle_{\tilde{\rho}_{\rm out}}
    ={\rm Real}(f_{01}^3)\hat{P}_2{\rm tr}\left(O\hat{\mathcal{E}}^{(2)}(\rho)\right)=\eta_2{\rm tr}\left(O\hat{\mathcal{E}}^{(2)}(\rho)\right)
\end{aligned},
\end{equation}
here we define $\eta_m:={\rm Real}(f_{01}^3)\hat{P}_m$.
By setting $O=I_{2^n}$, the expectation value of the control qubit measurement is
\begin{equation}\label{eq:ctrl_x}
\begin{aligned}
    \langle X\otimes I_{2^n}\rangle_{\tilde{\rho}_{\rm out}}=\eta_2
\end{aligned}.
\end{equation}
Furthermore, the above analysis can be easily generalized to the $m$th-order VCP, with $\langle X\otimes O\rangle_{\tilde{\rho}_{\rm out}}=\eta_m{\rm tr}\left(O\hat{\mathcal{E}}^{(m)}(\rho)\right)$ and $\langle X\otimes I_{2^n}\rangle_{\tilde{\rho}_{\rm out}}=\eta_m$.

Notice that without the existence of $\mathcal{F}$, we have ${\rm Real}(f_{01}^3)=1$. In contrast, the noise in the control qubit usually results in ${\rm Real}(f_{01}^3)<1$. However, VCP obtains results by dividing Eq.~\eqref{eq:ctrl_xo} by Eq.~\eqref{eq:ctrl_x}, thus canceling the coefficient ${\rm Real}(f_{01}^3)$, yielding the same value as the noise-free scenario.

\subsection{Impact of noise in the control subsystem of the VSP circuit}\label{apx:vsp_ctrl}

\begin{figure}[!ht]
    \centering
    \includegraphics[width=0.4\textwidth]{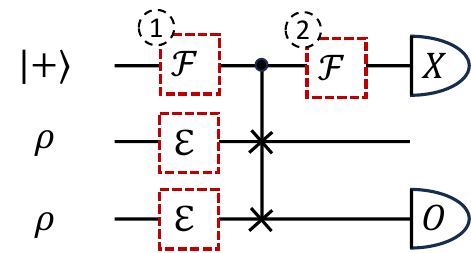}
    \caption{Tow locations where noise channel $\mathcal{F}$ can occur in the control subsystem of the VSP circuit.}\label{fig:ctrl_vsp}
\end{figure}

\begin{lemma}\label{lemma:ctrl}
    Suppose noise channel $\mathcal{F}=\sum_{i=0}^3q_i\overbracket{F_i}$ is a CPTP channel satisfies the properties
    \begin{equation}\nonumber
    \begin{aligned}
        \mathcal{F}(|i\rangle\langle j|)=\begin{cases}
            f_{ij}|i\rangle\langle j|, &i\neq j\\
            \sum_k f_{k}^{(i)}|k\rangle\langle k|, & i=j
        \end{cases}
    \end{aligned},
    \end{equation}
    where $f_{ij},f_k^{(i)}\in\mathbb{C}$.
    Then, for a local noise channel $\mathcal{F}$ and an $n$-qubit target quantum state $\rho$, it holds that
    \begin{equation}\nonumber
        \frac{\langle X\otimes O\rangle_{\tilde{\rho}_{\rm out}}}{\langle X\otimes I_{2^n}\rangle_{\tilde{\rho}_{\rm out}}}=\frac{\langle X\otimes O\rangle_{\rho_{\rm out}}}{\langle X\otimes I_{2^n}\rangle_{\rho_{\rm out}}},
    \end{equation}
    where $\tilde{\rho}_{\rm out}$ and $\rho_{\rm out}$ denote the output states of the virtual state purification circuit, with and without the existence of $\mathcal{F}$, respectively.
\end{lemma}
\textbf{Proof\ } Starting with the 2nd-order VSP circuit, let $\tilde{\rho}=\mathcal{E}(\rho)$ and $\sigma:=\mathcal{F}(|+\rangle\langle+|)$. The output state after the cSWAP gate is given by
\begin{equation}\nonumber
\begin{aligned}
    &\tilde{\rho}_{\rm out}=\overbracket{\rm cSWAP}\circ\left(\sigma\otimes \tilde{\rho}\otimes\tilde{\rho}\right)\\
    =&\left(\sigma_{00}\mathcal{F}(|0\rangle\langle0|)+\sigma_{11}\mathcal{F}(|1\rangle\langle1|)\right)\otimes\tilde{\rho}^{\otimes 2}\\
    &+\sigma_{01}f_{01}\ket{0}\bra{1}\otimes\tilde{\rho}^{\otimes 2}{\rm SWAP}+\sigma_{10}f_{10}\ket{1}\bra{0}\otimes {\rm SWAP}\tilde{\rho}^{\otimes 2}\\
    =&\left(\sigma_{00}\mathcal{F}(|0\rangle\langle0|)+\sigma_{11}\mathcal{F}(|1\rangle\langle1|)\right)\otimes\tilde{\rho}^{\otimes 2}\\
    &+\left(\sigma_{01}f_{01}\ket{0}\bra{1}+\sigma_{10}f_{10}\ket{1}\bra{0}\right)\otimes\sum_{ij}a_ia_j\ket{ij}\bra{ji}
\end{aligned},
\end{equation}
where $\tilde{\rho}=\sum_ia_i\ket{i}\bra{i}$ is the spectral decomposition of $\tilde{\rho}$. Then, we have
\begin{equation}\nonumber
\begin{aligned}
    \langle X\otimes O\rangle_{\tilde{\rho}_{\rm out}}&=\left(\sigma_{01}f_{01}+\sigma_{10}f_{10}\right)\sum_ia_i^2\tr(O\ket{i}\bra{i})\\
    &={\rm Real}(f_{01}^2)\tr(O\tilde{\rho}_{\rm out}^2)
\end{aligned},
\end{equation}
since $\sigma_{01}=\frac{1}{2}f_{01}=\sigma_{10}^*$. Similarly, by setting $O=I_{2^n}$, it holds that $\langle X\otimes I\rangle_{\tilde{\rho}_{\rm out}}={\rm Real}(f_{01}^2)\tr(\tilde{\rho}_{\rm out}^2)$. Analogously, for $m$th-order VSP, we have $\langle X\otimes O\rangle_{\tilde{\rho}_{\rm out}}={\rm Real}(f_{01}^2)\tr(O\tilde{\rho}_{\rm out}^m)$ and $\langle X\otimes I\rangle_{\tilde{\rho}_{\rm out}}={\rm Real}(f_{01}^2)\tr(\tilde{\rho}_{\rm out}^m)$. The coefficients introduced by $\mathcal{F}$ can thus be canceled through division, resulting in the same value as in the noise-free scenario. This completes the proof.

\subsection{Impact of noise in different locations of cSWAP gates}\label{apx:cswap_numerical}

Following the numerical setup described in~\ref{apx:comparison_vcp_vsp}, we now analyze the impact of various types of noise in the cSWAP gates within the single-layer VCP circuit numerically. The error rates for single-qubit, two-qubit, and cSWAP gates are assumed to be 0.001, 0.01, and 0.05, respectively. According to the categorization of noise types in the VCP circuit illustrated in Figure 3 in the main text, we consider three regions of noise affecting the cSWAP gates: \ding{172} noise in the control qubit of the cSWAP gates, \ding{173} noise in the last two subsystems between the two cSWAP layers, and \ding{175} noise in the target subsystem after the second cSWAP layer of the VCP circuit. We ignore the noise in \ding{174} as it does not affect the measurement expectation values. For each type of noise, we adjust the error rate from 0.05 to ten values evenly distributed from 0 to 0.05.

\begin{figure*}[t]
  \captionsetup[subfigure]{justification=centering, labelfont=bf}
  \centering
  \begin{subfigure}[t]{0.33\textwidth}
    \centering
    \includegraphics[width = \linewidth]{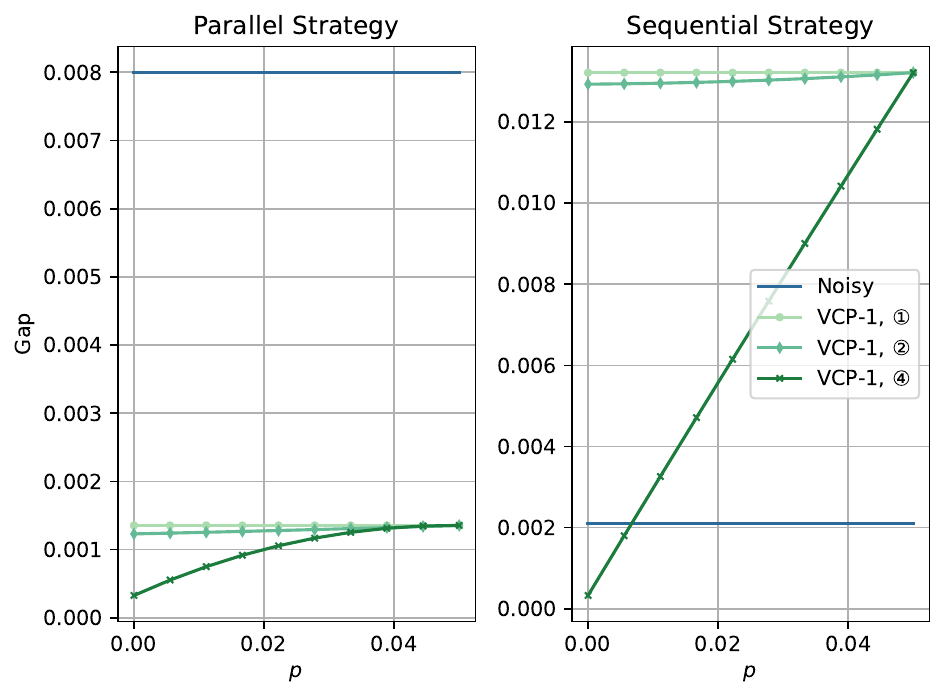}
    \subcaption{Depolarizing noise}
  \end{subfigure}%
  \hfill
  \begin{subfigure}[t]{0.33\textwidth}
    \centering
    \includegraphics[width = \linewidth]{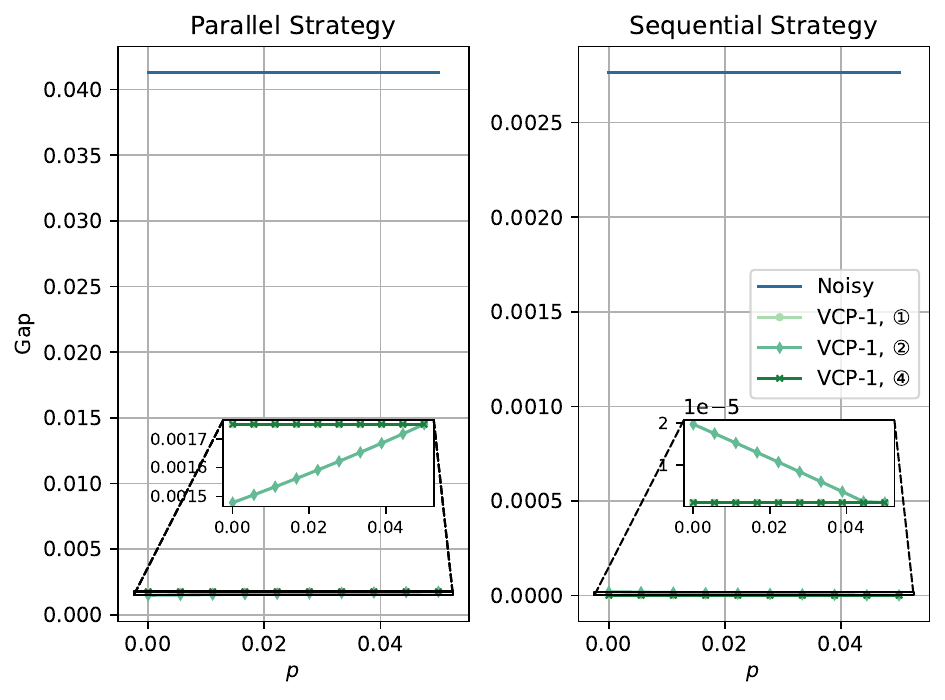}
    \subcaption{Dephasing noise}
  \end{subfigure}%
  \hfill
  \begin{subfigure}[t]{0.33\textwidth}
    \centering
    \includegraphics[width = \linewidth]{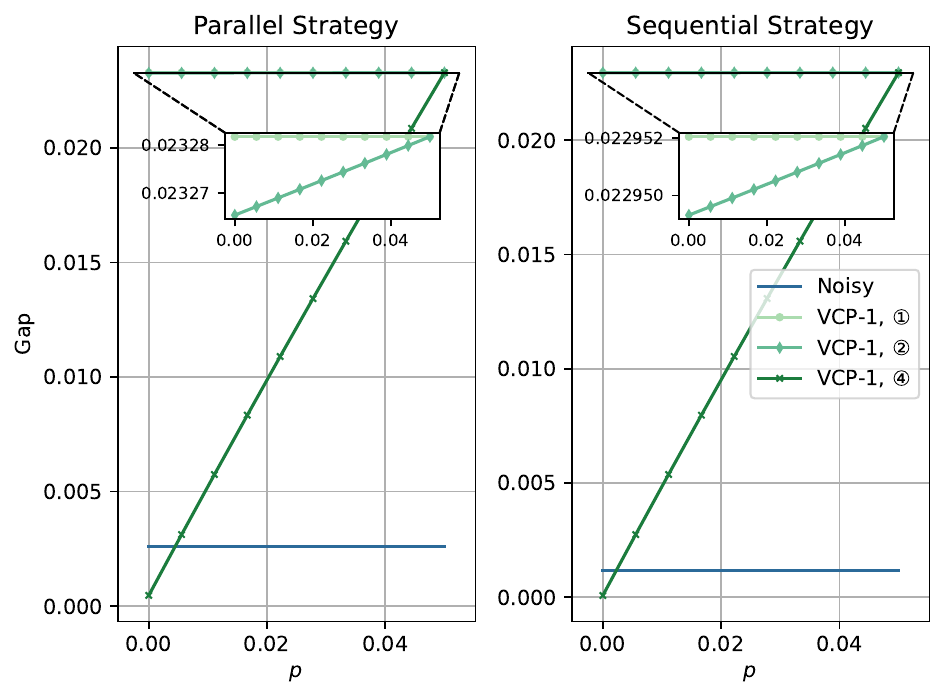}
    \subcaption{Amplitude damping noise}
  \end{subfigure}%

  \caption{Impacts of different types of noise in cSWAP gates for (a) depolarizing noise, (b) dephasing noise, and (c) amplitude damping noise. 
  The light green, green, and dark green lines indicate the parameter estimation gaps after applying single-layer VCP under different error rates $p$ for \ding{172} the control qubit of the cSWAP gate, \ding{173} the first cSWAP layer excluding the control qubit, and \ding{175} the target subsystem of the second cSWAP layer, respectively. The blue line depicts the gaps in the original noisy quantum circuits.}\label{fig:exp_cswap}
\end{figure*}

Set $N=5$ and $m=2$, Supplementary Figure~\ref{fig:exp_cswap} illustrates the corresponding numerical results, which align with the analysis above. First, the increasing error rate in \ding{172} does not affect the parameter estimation result, which introduces no systematic errors. Second, the gap under different error rates of noise in \ding{173} varies little, since it is suppressed automatically by VCP. As a comparison, noise in \ding{175} enlarges the gap significantly with the increasing error rate. For depolarizing noise and amplitude damping noise, we can observe the benefits of VCP when the error rate in \ding{175} is small. However, when the error rate becomes large, these advantages are diminished or may even lead to worse estimates, indicating the need for additional QEM methods to alleviate these effects. Dephasing noise is a special case since it does not affect the diagonal entries of the density matrix. As a result, the measurement outcome probabilities remain unchanged when measuring on a computational basis. Therefore, dephasing noise in \ding{175} will not introduce errors in the estimates.

\section{Variance of VCP-PEC circuits}\label{apx:variance}

In the VCP-PEC approach, the estimator is constructed by division as $\frac{\sum_i\alpha_i\langle X_i\rangle_{\tilde{\rho}_0}}{\sum_i \alpha_i\langle Y_i\rangle_{\tilde{\rho}_0}}$. In this formulation, the observables $X\otimes O$ and $X\otimes I_{2^n}$ are modified into $X_i$ and $Y_i$, respectively, to allow their expectation values to be taken on the same quantum state $\tilde{\rho}_0$. This is done such that $\langle X_i\rangle_{\tilde{\rho}_0}=\langle X\otimes O\rangle_{\tilde{\rho}_i}$ and $\langle Y_i\rangle_{\tilde{\rho}_0}=\langle X\otimes I_{2^n}\rangle_{\tilde{\rho}_i}$, where $\tilde{\rho}_i$ is the output state of the $i$-th VCP-PEC circuit. Suppose the nominator and the denominator are estimated using a number of $\nu$ circuit runs. Then, its variance can be calculated by\begin{equation}\label{eq:var_x/y}
\begin{aligned}
    {\rm Var}\left(\frac{x}{y}\right)\approx\frac{\mu_x^2}{\mu_y^2}\left(\frac{{\rm Var}(x)}{\mu_x^2}-2\frac{{\rm Cov}(x,y)}{\mu_x\mu_y}+\frac{{\rm Var}(y)}{\mu_y^2}\right)
\end{aligned},
\end{equation}
where $x$ and $y$ stands for the estimators of $\sum_i\alpha_i\langle X_i\rangle_{\tilde{\rho}_0}$ and $\sum_i \alpha_i\langle Y_i\rangle_{\tilde{\rho}_0}$, respectively, with expectation values $\mu_x=\eta_m{\rm tr}(O\hat{\mathcal{E}}^{(m)}(\rho))$ and $\mu_y=\eta_m$. 

To calculate the variance of VCP-PEC, we primarily focus on the scenario where the noise introduced by cSWAP gates satisfies the condition defined in Theorem 1. We start by considering the variance of VCP. The corresponding estimator is constructed as $\frac{\langle X\otimes O\rangle}{\langle X\otimes I_{2^n}\rangle}$, where the expectation values are taken from the output state of the VCP circuit. Let $x'$ and $y'$ be the estimators of $\langle X\otimes O\rangle$ and $\langle X\otimes I_{2^n}\rangle$, respectively. Follow the notations defined in Supplementary Note~\ref{apx:proof_thrm1}, and define the noise in both the ancillary and target subsystems introduced by cSWAP gates be $\mathcal{F'}$. Then, it can be easily derived that the expectation values of $x'$ and $y'$ are $\mu_{x'}=\eta_m\tr(O\mathcal{F'}\circ\hat{\mathcal{E}}^{(m)}(\rho))$ and $\mu_{y'}=\eta_m$, respectively.

Considering the case $m=2$, for the variance of $x'$, we have
\begin{equation}\nonumber
\begin{aligned}
    {\rm Var}(x')=\frac{1}{\nu}\left(\langle I_2\otimes O^2\rangle-\langle X\otimes O\rangle^2\right)
\end{aligned}.
\end{equation}
To estimate $\langle I_2\otimes O^2\rangle$, the impact of the last noise in the control subsystem can be ignored since $\mathcal{F}$ is trace-preserving. Then, it holds that 
\begin{equation}\label{eq:mu_io2}
\begin{aligned}
    \langle I_2\otimes O^2\rangle=&\langle I_2\otimes O^2\rangle_{\mathcal{I}_2\otimes\mathcal{F'}\otimes\mathcal{F'}\left(\overbracket{\rm cSWAP}(\tilde{\rho})\right)}\\
    =&\sigma_{00}f_0^{(0)}\langle0|I_2|0\rangle{\rm tr}(\mathcal{E'}(I_{2^n}/2^n)){\rm tr}(O^2\mathcal{E'}(\rho))\\
    &+\sigma_{00}f_1^{(0)}\langle1|I_2|1\rangle{\rm tr}(\mathcal{E'}(\rho)){\rm tr}(O^2\mathcal{E'}(I_{2^n}/2^n))\\
    &+\sigma_{11}f_0^{(1)}\langle0|I_2|0\rangle{\rm tr}(\mathcal{E'}(\rho)){\rm tr}(O^2\mathcal{E'}(I_{2^n}/2^n))\\
    &+\sigma_{11}f_1^{(1)}\langle1|I_2|1\rangle{\rm tr}(\mathcal{E'}(I_{2^n}/2^n)){\rm tr}(O^2\mathcal{E'}(\rho))\\
    =&\alpha{\rm tr}(O^2\mathcal{E'}(\rho))+\beta{\rm tr}(O^2\mathcal{E'}(I_{2^n}/2^n))
\end{aligned},
\end{equation}
where $\tilde{\rho}$ is defined in Eq.~\eqref{eq:ctrl12}, $\mathcal{E'}=\mathcal{F'}\circ\mathcal{E}$, $\alpha=\sigma_{00}f_0^{(0)}+\sigma_{11}f_1^{(1)}$ and $\beta=\sigma_{00}f_1^{(0)}+\sigma_{11}f_0^{(1)}$. Thus,
\begin{equation}\label{eq:var_x}
\begin{aligned}
    {\rm Var}(x')=\frac{1}{\nu}\Big\{\Big.\alpha{\rm tr}\left(O^2\mathcal{E'}(\rho)\right)+\beta{\rm tr}\left(O^2\mathcal{E'}(I_{2^n}/2^n)\right)-\eta_2^2{\rm tr}\left(O\mathcal{F'}\circ\hat{\mathcal{E}}^{(2)}(\rho)\right)^2\Big.\Big\}
\end{aligned}.
\end{equation}
Similarly, we have
\begin{equation}\label{eq:var_y}
\begin{aligned}
    {\rm Var}(y')&=\frac{1}{\nu}\left(\langle I_2\otimes I_{2^n}\rangle -\langle X\otimes I_{2^n}\rangle^2\right)=\frac{1}{\nu}\left(1-\eta_2^2\right)
\end{aligned}.
\end{equation}
Replacing $O^2$ with $O$ in Eq.~\eqref{eq:mu_io2}, the covariance turns out to be
\begin{equation}\label{eq:cov_xy}
\begin{aligned}
    {\rm Cov}(x',y')&=\frac{1}{\nu}\left\{\langle I_2\otimes O\rangle-\langle X\otimes O\rangle\langle X\otimes I_{2^n}\rangle\right\}\\
    &\begin{aligned}
        =\frac{1}{\nu}\Big\{\Big.\alpha{\rm tr}\left(O\mathcal{E'}(\rho)\right)+\beta{\rm tr}\left(O\mathcal{E'}(I_{2^n}/2^n)\right)-\eta_2^2{\rm tr}\left(O\mathcal{F'}\circ\hat{\mathcal{E}}^{(2)}(\rho)\right)\Big.\Big\}
    \end{aligned}
\end{aligned}.
\end{equation}
Substituting Eqs.~\eqref{eq:var_x},~\eqref{eq:var_y} and~\eqref{eq:cov_xy} into Eq.~\eqref{eq:var_x/y}, the final variance reads
\begin{equation}
\begin{aligned}
    {\rm Var}\left(\frac{x'}{y'}\right)\approx&\frac{1}{\nu\eta_2^2}\Big\{\Big.\alpha{\rm tr}\left(O^2\mathcal{E'}(\rho)\right)+\beta{\rm tr}\left(O^2\mathcal{E'}(I_{2^n}/2^n)\right)\\
    &-2\left(\alpha{\rm tr}\left(O\mathcal{E'}(\rho)\right)+\beta{\rm tr}\left(O\mathcal{E'}(I_{2^n}/2^n)\right)\right){\rm tr}\left(O\mathcal{F'}\circ\hat{\mathcal{E}}^{(2)}(\rho)\right)\\
    &
    +{\rm tr}\left(O\mathcal{F'}\circ\hat{\mathcal{E}}^{(2)}(\rho)\right)^2\Big.\Big\}\\
    \le&\frac{1}{\nu\eta_2^2}\left\{\Vert O^2\Vert_{\infty}+3\Vert O\Vert_{\infty}^2\right\}
\end{aligned},
\end{equation}
where $\Vert\cdot\Vert_{\infty}$ denotes the spectral norm.
Here, we assume $\eta^2$ is small since it exponentially decreases with the number of noise. The last inequality utilizes the fact that $\alpha,\beta\ge0$ and $\alpha+\beta=1$ since $\mathcal{F}$ is a CPTP channel. 

When considering VCP-PEC, suppose the inverse noise operation $\mathcal{F'}^{-1}$ is decomposed with coefficients $\alpha_i$, where $\sum_i\alpha_i=1$ and $\gamma=\sum_i|\alpha_i|\ge1$. According to the analysis of PEC, it holds that ${\rm Var}(x)={\mathcal{O}}(\gamma^2{\rm Var}(x'))$~\cite{PhysRevLett.119.180509,PhysRevX.8.031027}. Besides, since PEC is applied only to the target subsystem and does not affect the measurement of the observable $X\otimes I_{2^n}$, we have ${\rm Var}(y)={\rm Var}(y')$. For the covariance, it is given that ${\rm Cov}(x,y)=\sum_{i,j}\alpha_i\alpha_j{\rm Cov}(x_i, y_j)$, where $x_i$ and $y_j$ are the estimators for $\langle X_i\rangle_{\tilde{\rho}_0}$ and $\langle Y_j\rangle_{\tilde{\rho}_0}$, respectively. In line with the assumptions from Refs.~\cite{RevModPhys.95.045005,cai2023practicalframeworkquantumerror}, we assume ${\rm Cov}(x_i,y_j)\sim{\rm Cov}(x',y')$ since the ensemble of circuits induced by PEC are variants of the primary circuit. Thus, the covariance can be simplified to ${\rm Cov}(x,y)=\sum_{i,j}\alpha_i\alpha_j{\rm Cov}(x', y')={\rm Cov}(x', y')$, given that $\sum_i\alpha_i=1$. As a consequence,
\begin{equation}\label{eq:var_x/y_final}
\begin{aligned}
    {\rm Var}\left(\frac{x}{y}\right)\approx&\frac{\mu_x^2}{\mu_y^2}\left(\frac{\gamma^2{\rm Var}(x')}{\mu_x^2}-2\frac{{\rm Cov}(x',y')}{\mu_x\mu_y}+\frac{{\rm Var}(y')}{\mu_y^2}\right) \\
    \le&\frac{1}{\nu\eta_2^2}\Big\{\Big.\gamma^2\Vert O^2\Vert_{\infty}+3\Vert O\Vert_{\infty}^2+\gamma^2\eta_2^2\tr\left(O\mathcal{F'}\circ\hat{\mathcal{E}}^{(2)}(\rho)\right)^2\\
    &\quad\quad+2\eta_2^2\tr\left(O\mathcal{F'}\circ\hat{\mathcal{E}}^{(2)}(\rho)\right)\tr\left(O\hat{\mathcal{E}}^{(2)}(\rho)\right)-\eta_2^2\tr\left(O\hat{\mathcal{E}}^{(2)}(\rho)\right)^2\Big.\Big\}\\
    \le&\frac{1}{\nu\eta_2^2}\left\{\gamma^2\Vert O^2\Vert_{\infty}+(\gamma^2\eta_2^2+2\eta_2^2+3)\Vert O\Vert_{\infty}^2\right\}
\end{aligned}.
\end{equation}

Considering a bounded observable $O$, to limit the variation to be $\epsilon^2$, the sampling cost scales as $\nu={\mathcal{O}}\left(\frac{\gamma^2}{\epsilon^2\eta_2^2}\right)$. For the case where $m\ge2$, it can be verified that the sampling cost similarly scales as $\nu=O\left(\frac{\gamma^2}{\epsilon^2\eta_m^2}\right)$. Additionally, since $O$ is a bounded observable, a similar result can be obtained even without the assumption that ${\rm Cov}(x_i, y_j)\sim{\rm Cov}(x', y')$.

\section{Comparative analysis of sampling costs for noise in the control subsystem of VCP circuits}\label{apx:sampling_comparison}

The analysis of variance in~\ref{apx:variance} indicates that the statistical error introduced by each noise in the control subsystem is characterized by ${\rm Real}(f_{01})^{-2}$. It is interesting to compare the sampling cost amplified by the noise in the control qubit with the sampling cost of introducing other QEM protocols to mitigate this noise.

\textbf{Case 1.} Let $\mathcal{F}$ be a dephasing channel, with noise level $p\in[0,1]$, i.e.,
\begin{equation}\nonumber
    \mathcal{F}(\rho)=(1-p)\rho+pZ\rho Z.
\end{equation}
Then, we have ${\rm Real}(f_{01})=1-2p$. Hence, the additional sampling cost introduced by each dephasing noise equals $(1-2p)^{-2}$, which matches the optimal cost for mitigating it~\cite{PhysRevResearch.3.033178,Jiang2021physical}.

\textbf{Case 2.} Let $\mathcal{F}$ be a depolarizing channel, with noise level $p\in[0,1]$, that maps a quantum state $\rho$ into
\begin{equation}\nonumber
    \mathcal{F}(\rho)=(1-p)\rho+\frac{p}{2}I_2.
\end{equation}
It can be easily found that ${\rm Real}(f_{01})=1-p$, which indicates that the additional sampling cost introduced by each depolarizing noise is $(1-p)^{-2}$. As a comparison, it has been proved that the optimal cost to mitigate depolarizing noise requires $\left(\frac{1+p/2}{1-p}\right)^{2}$~\cite{PhysRevResearch.3.033178,Jiang2021physical}, which is clearly greater than simply ignoring the noise.

\textbf{Case 3.} Let $\mathcal{F}$ be a amplitude damping channel, with Kraus operators $F_0=|0\rangle\langle 0|+\sqrt{1-p}|1\rangle\langle1|$ and $F_1=\sqrt{p}|0\rangle\langle 1|$, that maps a quantum state $\rho$ into
\begin{equation}\nonumber
    \mathcal{F}(\rho)=F_0\rho F_0^{\dagger}+F_1\rho F_1^{\dagger},
\end{equation}
where $p\in[0,1]$.
Then, it holds that ${\rm Real}(f_{01})=\sqrt{1-p}$, indicating that the sampling cost is amplified by a factor of $(1-p)^{-1}$ for each instance of amplitude damping noise. In contrast, the minimal cost to mitigate the amplitude damping noise is shown to be $\left(\frac{1+p}{1-p}\right)^{2}$~\cite{Jiang2021physical}. Once again, it is more efficient to do nothing regarding this type of noise in the control subsystem.

These results are reasonable because the sampling cost for QEM we mentioned above is designed to handle arbitrary circuits, whereas the sampling cost ${\rm Real}(f_{01})^{-2}$ applies specifically to virtual purification-based circuits. It can be easily verified that the comparisons in these cases also hold for VSP circuits. Nonetheless, the properties we have discovered are valuable for designing an efficient framework to enhance the efficacy of these virtual purification methods.

\section{Numerical simulation: single-parameter estimation}\label{apx:numerical_single}

In the context of the Zeeman Hamiltonian discussed in~\ref{apx:comparison_vcp_vsp}, we evaluate the efficacy of five methods: noisy method, VSP, VCP, VSP-PEC and VCP-PEC. In specific, a sequential scheme is considered, where the Hamiltonian $H=\lambda Z/2$ is applied $N$ times in sequence. Let the local parameter $\lambda=\frac{\pi}{4}\times 10^{-4}$, and $N\in\{10, 50, 100, 200, 500, 800, 1,000\}$. Suppose the error rates for single-qubit gates and cSWAP gates are 0.001 and 0.05, respectively. Particularly for VCP and VCP-PEC, a layer-wise implementation is employed with a maximum of 5 layers, and the following results present their performance at the optimal layer $L^*$.

\begin{figure*}[!ht]
  \captionsetup[subfigure]{justification=centering, labelfont=bf}
  \centering
  \begin{subfigure}[!ht]{0.33\textwidth}
    \centering
    \includegraphics[width = \linewidth]{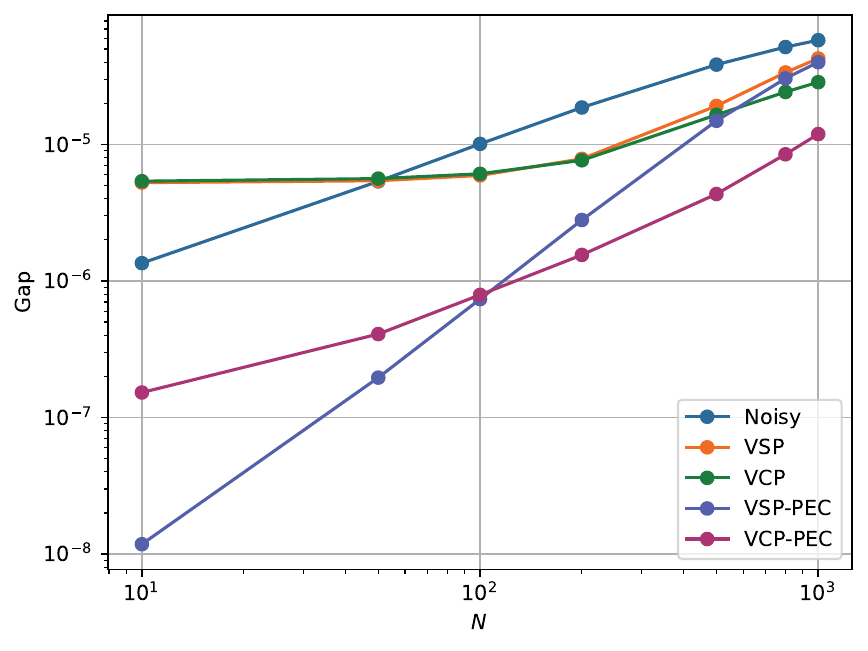}
    \subcaption{Depolarizing noise}
  \end{subfigure}
  \hfill
  \begin{subfigure}[!ht]{0.33\textwidth}
    \centering
    \includegraphics[width = \linewidth]{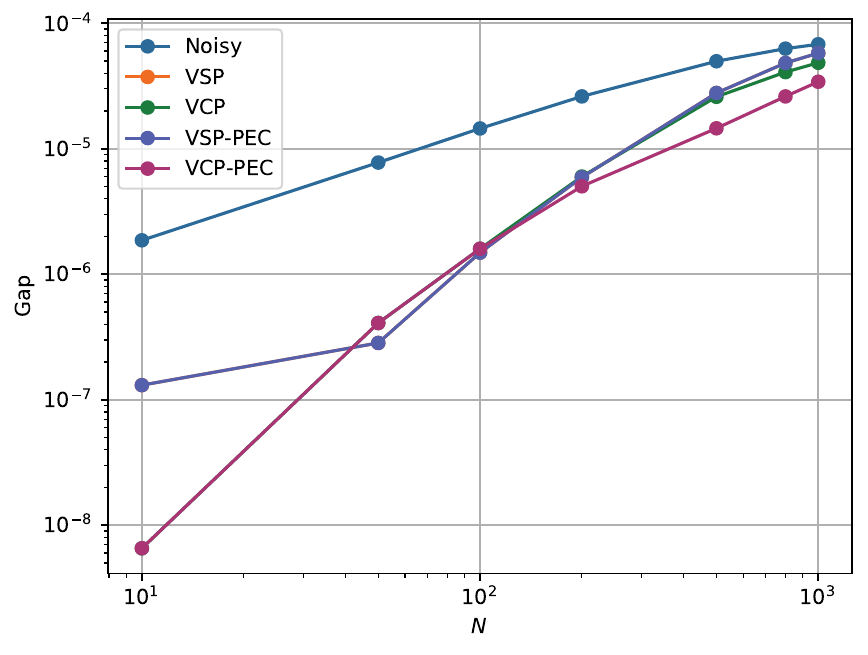}
    \subcaption{Dephasing noise}
  \end{subfigure}%
  \hfill
  \begin{subfigure}[!ht]{0.33\textwidth}
    \centering
    \includegraphics[width = \linewidth]{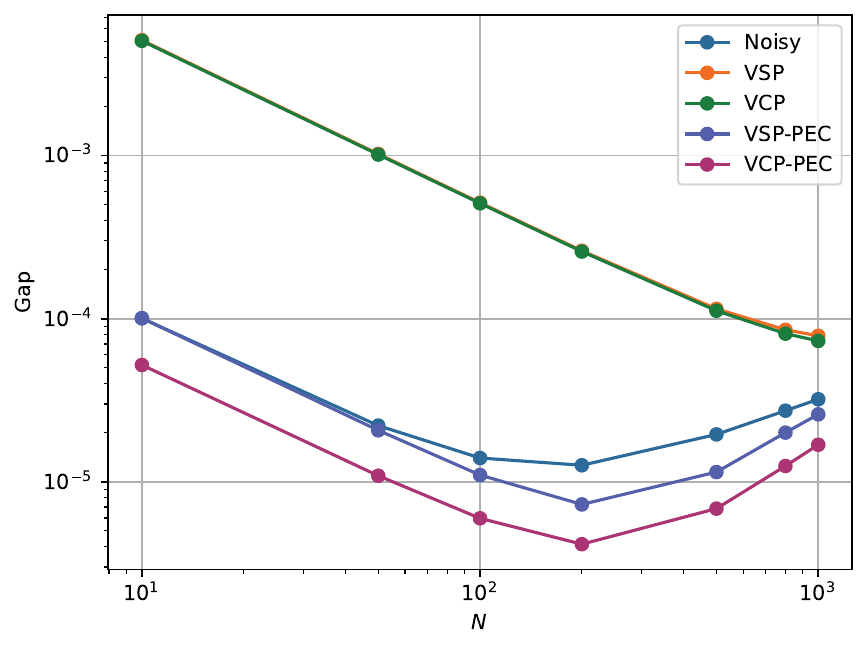}
    \subcaption{Amplitude damping noise}
  \end{subfigure}

  \vspace{\baselineskip}
  
  \begin{subfigure}[!ht]{0.33\textwidth}
    \centering
    \includegraphics[width = \linewidth]{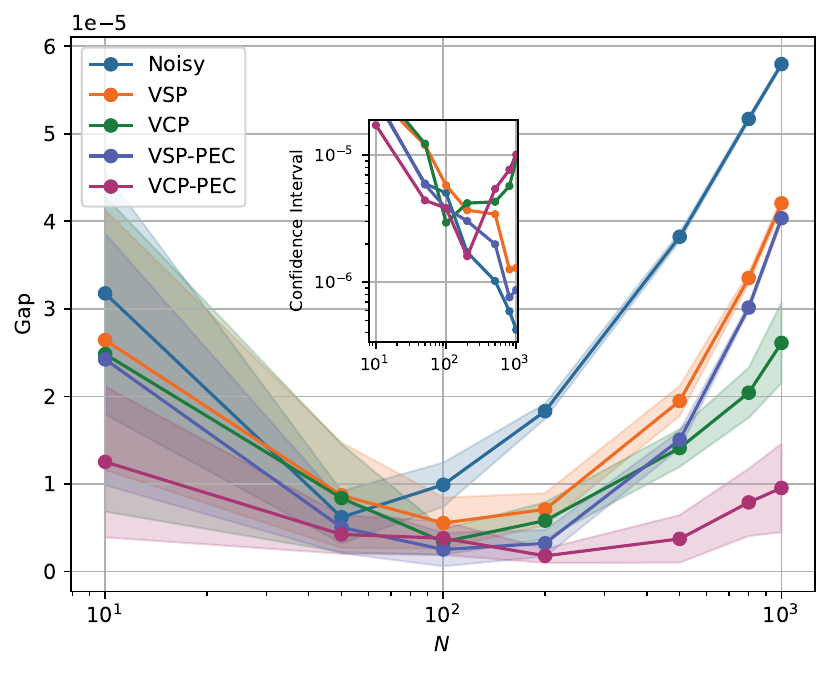}
    \subcaption{Depolarizing noise, $\nu=10^7$}
  \end{subfigure}
  \hfill
  \begin{subfigure}[!ht]{0.325\textwidth}
    \centering
    \includegraphics[width = \linewidth]{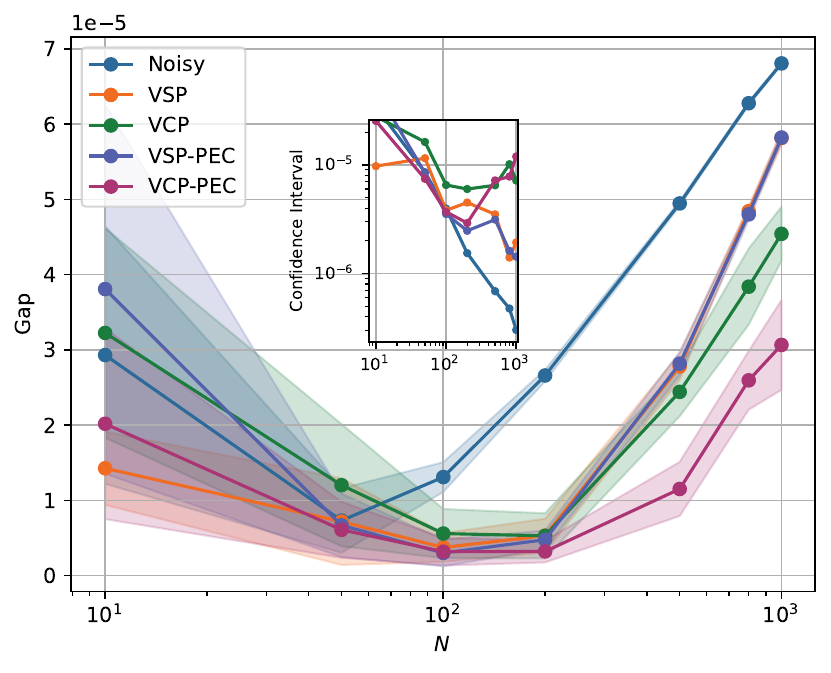}
    \subcaption{Dephasing noise, $\nu=10^7$}
  \end{subfigure}
  \hfill
  \begin{subfigure}[!ht]{0.33\textwidth}
    \centering
    \includegraphics[width = \linewidth]{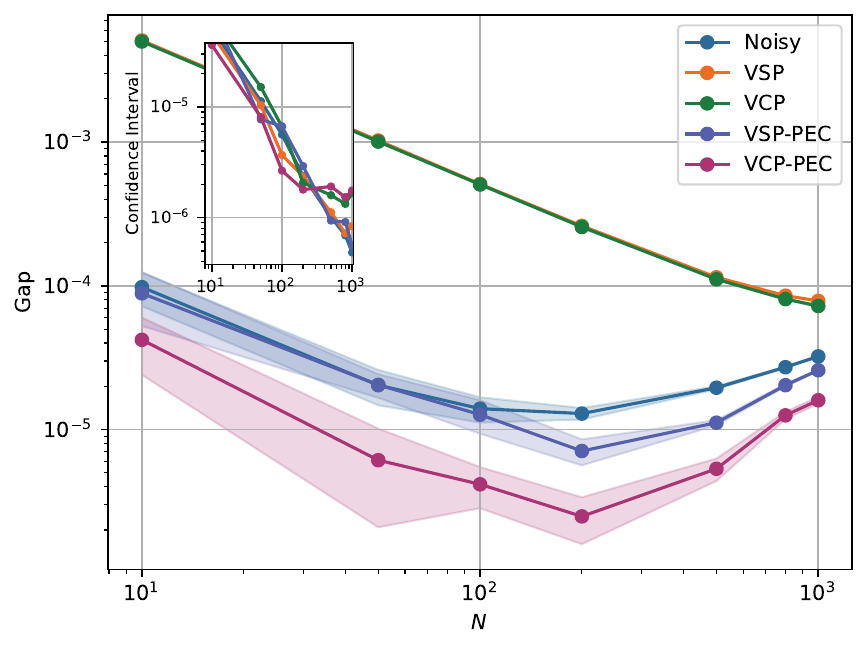}
    \subcaption{Amplitude damping noise, $\nu=10^7$}
  \end{subfigure}
    
  \caption{Single-parameter estimation gaps for different $N$ under different types of noise channels. (a)-(c) plot the performance of different methods with infinite measurement shots, while (d)-(f) depict those with measurement shots $\nu=10^7$. In scenarios with a limited number of measurement shots, experiments are conducted 10 times to calculate the mean values of the gaps and the 95\% confidence intervals. These are represented by solid lines and shaded areas, respectively. Additionally, the exact values of these confidence intervals are provided in the inset subfigures.}\label{fig:exp_seq}
\end{figure*}

Supplementary Figures~\ref{fig:exp_seq}(a)-(c) illustrate the corresponding results with an infinite number of measurement shots. Without the assistance of PEC, the original virtual purification methods offer limited advantages over the performance of the original noisy circuit and may even exhibit worse behavior in some instances. However, the introduction of PEC significantly enhances the performance of both virtual purification methods. This aligns with theoretical analysis, which suggests that VSP-PEC performs well when the accumulated error is low (i.e., when $N$ is small), whereas VCP-PEC generally exhibits a clear advantage over VSP-PEC across most scenarios. With an infinite number of measurement shots, the optimal layers $L^*$ for VCP under different values of $N$ are generally 1 or 2, while VCP-PEC achieves optimal performance with 5 layers in most cases. Thus, by incorporating PEC, additional VCP layers can be utilized to further suppress errors.

In addition, we assess the behavior of these methods with a limited number of measurement shots, specifically $\nu=10^7$. By repeating the experiments 10 times, Supplementary Figures~\ref{fig:exp_seq}(d)-(f) display the mean values of the gaps between the estimated and exact values for each method, illustrated by solid lines. The shaded areas represent the corresponding 95\% confidence intervals, with the exact values of these intervals detailed in the inset subfigure. Similar to the multi-parameter estimation tasks represented in the main text, the error-mitigated estimators are more sensitive to variations in the sampled measurement outcomes, and all four QEM methods exhibit wider confidence intervals than the original noisy quantum circuits. For example, when $N=1,000$, VCP and VCP-PEC display wider confidence intervals compared to VSP and VSP-PEC. However, the introduction of PEC does not significantly increase the confidence interval. Moreover, the range of gaps for VCP-PEC is generally smaller than those of the other methods, underscoring its practical effectiveness. Furthermore, the optimal number of layers $L^*$ for VCP typically remains 1 or 2, while for VCP-PEC, it is usually reduced to 2 or 3 under a limited number of measurement shots, as additional VCP layers tend to increase variance.

In quantum metrology, it is generally expected that the gap between estimated and true values decreases as $N$ increases. However, in this scenario, higher precision is achieved only when $N$ is roughly less than 100. Beyond this point, the presence of noise diminishes the benefit of repeated use of the encoding channel $\mathcal{U}_{\lambda}$. Nevertheless, when $\lambda$ approaches zero, which corresponds to a larger Fisher information, our additional numerical simulations indicate that the gap indeed decreases with increasing $N$. This observation also highlights the advantage of the sequential feedback scheme.

\bibliography{ref}

@article{PhysRevLett.96.010401,
  title = {Quantum Metrology},
  author = {Giovannetti, Vittorio and Lloyd, Seth and Maccone, Lorenzo},
  journal = {Phys. Rev. Lett.},
  volume = {96},
  issue = {1},
  pages = {010401},
  numpages = {4},
  year = {2006},
  month = {Jan},
  publisher = {American Physical Society},
  doi = {10.1103/PhysRevLett.96.010401},
  url = {https://link.aps.org/doi/10.1103/PhysRevLett.96.010401}
}

@article{maccone2021,
	author = {Giovannetti, Vittorio and Lloyd, Seth and Maccone, Lorenzo},
	date = {2011/04/01},
	doi = {10.1038/nphoton.2011.35},
	id = {Giovannetti2011},
	isbn = {1749-4893},
	journal = {Nature Photonics},
	number = {4},
	pages = {222--229},
	title = {Advances in quantum metrology},
	url = {https://doi.org/10.1038/nphoton.2011.35},
	volume = {5},
	year = {2011}}

@article{RevModPhys.89.035002,
  title = {Quantum sensing},
  author = {Degen, C. L. and Reinhard, F. and Cappellaro, P.},
  journal = {Rev. Mod. Phys.},
  volume = {89},
  issue = {3},
  pages = {035002},
  numpages = {39},
  year = {2017},
  month = {Jul},
  publisher = {American Physical Society},
  doi = {10.1103/RevModPhys.89.035002},
  url = {https://link.aps.org/doi/10.1103/RevModPhys.89.035002}
}

@article{RevModPhys.90.035005,
  title = {Quantum metrology with nonclassical states of atomic ensembles},
  author = {Pezz\`e, Luca and Smerzi, Augusto and Oberthaler, Markus K. and Schmied, Roman and Treutlein, Philipp},
  journal = {Rev. Mod. Phys.},
  volume = {90},
  issue = {3},
  pages = {035005},
  numpages = {70},
  year = {2018},
  month = {Sep},
  publisher = {American Physical Society},
  doi = {10.1103/RevModPhys.90.035005},
  url = {https://link.aps.org/doi/10.1103/RevModPhys.90.035005}
}

@article{PhysRevA.88.022318,
  title = {Quantum computations without definite causal structure},
  author = {Chiribella, Giulio and D'Ariano, Giacomo Mauro and Perinotti, Paolo and Valiron, Benoit},
  journal = {Phys. Rev. A},
  volume = {88},
  issue = {2},
  pages = {022318},
  numpages = {15},
  year = {2013},
  month = {Aug},
  publisher = {American Physical Society},
  doi = {10.1103/PhysRevA.88.022318},
  url = {https://link.aps.org/doi/10.1103/PhysRevA.88.022318}
}

@article{PhysRevLett.124.190503,
  title = {Quantum Metrology with Indefinite Causal Order},
  author = {Zhao, Xiaobin and Yang, Yuxiang and Chiribella, Giulio},
  journal = {Phys. Rev. Lett.},
  volume = {124},
  issue = {19},
  pages = {190503},
  numpages = {6},
  year = {2020},
  month = {May},
  publisher = {American Physical Society},
  doi = {10.1103/PhysRevLett.124.190503},
  url = {https://link.aps.org/doi/10.1103/PhysRevLett.124.190503}
}

@article{PhysRevLett.123.210502,
  title = {Reversing Unknown Quantum Transformations: Universal Quantum Circuit for Inverting General Unitary Operations},
  author = {Quintino, Marco T\'ulio and Dong, Qingxiuxiong and Shimbo, Atsushi and Soeda, Akihito and Murao, Mio},
  journal = {Phys. Rev. Lett.},
  volume = {123},
  issue = {21},
  pages = {210502},
  numpages = {5},
  year = {2019},
  month = {Nov},
  publisher = {American Physical Society},
  doi = {10.1103/PhysRevLett.123.210502},
  url = {https://link.aps.org/doi/10.1103/PhysRevLett.123.210502}
}

@article{PhysRevLett.127.200504,
  title = {Strict Hierarchy between Parallel, Sequential, and Indefinite-Causal-Order Strategies for Channel Discrimination},
  author = {Bavaresco, Jessica and Murao, Mio and Quintino, Marco T\'ulio},
  journal = {Phys. Rev. Lett.},
  volume = {127},
  issue = {20},
  pages = {200504},
  numpages = {7},
  year = {2021},
  month = {Nov},
  publisher = {American Physical Society},
  doi = {10.1103/PhysRevLett.127.200504},
  url = {https://link.aps.org/doi/10.1103/PhysRevLett.127.200504}
}

@article{Davidovich2011,
	author = {Escher, B. M. and de Matos Filho, R. L. and Davidovich, L.},
	date = {2011/05/01},
	doi = {10.1038/nphys1958},
	id = {Escher2011},
	isbn = {1745-2481},
	journal = {Nature Physics},
	number = {5},
	pages = {406--411},
	title = {General framework for estimating the ultimate precision limit in noisy quantum-enhanced metrology},
	url = {https://doi.org/10.1038/nphys1958},
	volume = {7},
	year = {2011}}

@article{Madalin2012,
	author = {Demkowicz-Dobrza{\'n}ski, Rafa{\l} and Ko{\l}ody{\'n}ski, Jan and Gu{\c t}{\u a}, M{\u a}d{\u a}lin},
	date = {2012/09/18},
	doi = {10.1038/ncomms2067},
	id = {Demkowicz-Dobrza{\'n}ski2012},
	isbn = {2041-1723},
	journal = {Nature Communications},
	number = {1},
	pages = {1063},
	title = {The elusive Heisenberg limit in quantum-enhanced metrology},
	url = {https://doi.org/10.1038/ncomms2067},
	volume = {3},
	year = {2012}}

@article{Dobrzanski18,
    url = {https://doi.org/10.1515/qmetro-2018-0002},
    title = {Precision Limits in Quantum Metrology with Open Quantum Systems},
    title = {},
    author = {J. F. Haase and A. Smirne and S. F. Huelga and J. Kołodynski and R. Demkowicz-Dobrzanski},
    pages = {13--39},
    volume = {5},
    number = {1},
    journal = {Quantum Measurements and Quantum Metrology},
    doi = {doi:10.1515/qmetro-2018-0002},
    year = {2016},
    lastchecked = {2024-05-29}
}

@article{PhysRevLett.112.150802,
  title = {Quantum Error Correction for Metrology},
  author = {Kessler, E. M. and Lovchinsky, I. and Sushkov, A. O. and Lukin, M. D.},
  journal = {Phys. Rev. Lett.},
  volume = {112},
  issue = {15},
  pages = {150802},
  numpages = {5},
  year = {2014},
  month = {Apr},
  publisher = {American Physical Society},
  doi = {10.1103/PhysRevLett.112.150802},
  url = {https://link.aps.org/doi/10.1103/PhysRevLett.112.150802}
}

@article{PhysRevLett.112.080801,
  title = {Improved Quantum Metrology Using Quantum Error Correction},
  author = {D\"ur, W. and Skotiniotis, M. and Fr\"owis, F. and Kraus, B.},
  journal = {Phys. Rev. Lett.},
  volume = {112},
  issue = {8},
  pages = {080801},
  numpages = {5},
  year = {2014},
  month = {Feb},
  publisher = {American Physical Society},
  doi = {10.1103/PhysRevLett.112.080801},
  url = {https://link.aps.org/doi/10.1103/PhysRevLett.112.080801}
}

@article{robust2015,
	author = {Lu, Xiao-Ming and Yu, Sixia and Oh, C. H.},
	date = {2015/06/08},
	doi = {10.1038/ncomms8282},
	id = {Lu2015},
	isbn = {2041-1723},
	journal = {Nature Communications},
	number = {1},
	pages = {7282},
	title = {Robust quantum metrological schemes based on protection of quantum Fisher information},
	url = {https://doi.org/10.1038/ncomms8282},
	volume = {6},
	year = {2015}}

@article{PhysRevLett.115.200501,
  title = {Quantum Error-Correction-Enhanced Magnetometer Overcoming the Limit Imposed by Relaxation},
  author = {Herrera-Mart\'{\i}, David A. and Gefen, Tuvia and Aharonov, Dorit and Katz, Nadav and Retzker, Alex},
  journal = {Phys. Rev. Lett.},
  volume = {115},
  issue = {20},
  pages = {200501},
  numpages = {5},
  year = {2015},
  month = {Nov},
  publisher = {American Physical Society},
  doi = {10.1103/PhysRevLett.115.200501},
  url = {https://link.aps.org/doi/10.1103/PhysRevLett.115.200501}
}

@article{PhysRevLett.122.040502,
  title = {Ancilla-Free Quantum Error Correction Codes for Quantum Metrology},
  author = {Layden, David and Zhou, Sisi and Cappellaro, Paola and Jiang, Liang},
  journal = {Phys. Rev. Lett.},
  volume = {122},
  issue = {4},
  pages = {040502},
  numpages = {6},
  year = {2019},
  month = {Jan},
  publisher = {American Physical Society},
  doi = {10.1103/PhysRevLett.122.040502},
  url = {https://link.aps.org/doi/10.1103/PhysRevLett.122.040502}
}

@article{Liang2018,
	author = {Zhou, Sisi and Zhang, Mengzhen and Preskill, John and Jiang, Liang},
	date = {2018/01/08},
	doi = {10.1038/s41467-017-02510-3},
	id = {Zhou2018},
	isbn = {2041-1723},
	journal = {Nature Communications},
	number = {1},
	pages = {78},
	title = {Achieving the Heisenberg limit in quantum metrology using quantum error correction},
	url = {https://doi.org/10.1038/s41467-017-02510-3},
	volume = {9},
	year = {2018}}

@article{RevModPhys.95.045005,
  title = {Quantum error mitigation},
  author = {Cai, Zhenyu and Babbush, Ryan and Benjamin, Simon C. and Endo, Suguru and Huggins, William J. and Li, Ying and McClean, Jarrod R. and O'Brien, Thomas E.},
  journal = {Rev. Mod. Phys.},
  volume = {95},
  issue = {4},
  pages = {045005},
  numpages = {37},
  year = {2023},
  month = {Dec},
  publisher = {American Physical Society},
  doi = {10.1103/RevModPhys.95.045005},
  url = {https://link.aps.org/doi/10.1103/RevModPhys.95.045005}
}

@article{PhysRevLett.119.180509,
  title = {Error Mitigation for Short-Depth Quantum Circuits},
  author = {Temme, Kristan and Bravyi, Sergey and Gambetta, Jay M.},
  journal = {Phys. Rev. Lett.},
  volume = {119},
  issue = {18},
  pages = {180509},
  numpages = {5},
  year = {2017},
  month = {Nov},
  publisher = {American Physical Society},
  doi = {10.1103/PhysRevLett.119.180509},
  url = {https://link.aps.org/doi/10.1103/PhysRevLett.119.180509}
}

@article{PhysRevX.8.031027,
  title = {Practical Quantum Error Mitigation for Near-Future Applications},
  author = {Endo, Suguru and Benjamin, Simon C. and Li, Ying},
  journal = {Phys. Rev. X},
  volume = {8},
  issue = {3},
  pages = {031027},
  numpages = {21},
  year = {2018},
  month = {Jul},
  publisher = {American Physical Society},
  doi = {10.1103/PhysRevX.8.031027},
  url = {https://link.aps.org/doi/10.1103/PhysRevX.8.031027}
}

@article{PhysRevResearch.3.033098,
  title = {Unified approach to data-driven quantum error mitigation},
  author = {Lowe, Angus and Gordon, Max Hunter and Czarnik, Piotr and Arrasmith, Andrew and Coles, Patrick J. and Cincio, Lukasz},
  journal = {Phys. Rev. Res.},
  volume = {3},
  issue = {3},
  pages = {033098},
  numpages = {12},
  year = {2021},
  month = {Jul},
  publisher = {American Physical Society},
  doi = {10.1103/PhysRevResearch.3.033098},
  url = {https://link.aps.org/doi/10.1103/PhysRevResearch.3.033098}
}

@article{PhysRevX.11.041036,
  title = {Virtual Distillation for Quantum Error Mitigation},
  author = {Huggins, William J. and McArdle, Sam and O'Brien, Thomas E. and Lee, Joonho and Rubin, Nicholas C. and Boixo, Sergio and Whaley, K. Birgitta and Babbush, Ryan and McClean, Jarrod R.},
  journal = {Phys. Rev. X},
  volume = {11},
  issue = {4},
  pages = {041036},
  numpages = {25},
  year = {2021},
  month = {Nov},
  publisher = {American Physical Society},
  doi = {10.1103/PhysRevX.11.041036},
  url = {https://link.aps.org/doi/10.1103/PhysRevX.11.041036}
}

@article{PhysRevLett.129.250503,
  title = {Error-Mitigated Quantum Metrology via Virtual Purification},
  author = {Yamamoto, Kaoru and Endo, Suguru and Hakoshima, Hideaki and Matsuzaki, Yuichiro and Tokunaga, Yuuki},
  journal = {Phys. Rev. Lett.},
  volume = {129},
  issue = {25},
  pages = {250503},
  numpages = {6},
  year = {2022},
  month = {Dec},
  publisher = {American Physical Society},
  doi = {10.1103/PhysRevLett.129.250503},
  url = {https://link.aps.org/doi/10.1103/PhysRevLett.129.250503}
}

@article{hama2023quantum,
  title={Quantum-Error-Mitigation Circuit Groups for Noisy Quantum Metrology},
  author={Hama, Yusuke and Nishi, Hirofumi},
  journal={arXiv preprint arXiv:2303.01820},
  year={2023}
}

@article{PhysRevX.11.031057,
  title = {Exponential Error Suppression for Near-Term Quantum Devices},
  author = {Koczor, B\'alint},
  journal = {Phys. Rev. X},
  volume = {11},
  issue = {3},
  pages = {031057},
  numpages = {30},
  year = {2021},
  month = {Sep},
  publisher = {American Physical Society},
  doi = {10.1103/PhysRevX.11.031057},
  url = {https://link.aps.org/doi/10.1103/PhysRevX.11.031057}
}

@article{Jiang2021physical,
  doi = {10.22331/q-2021-12-07-600},
  url = {https://doi.org/10.22331/q-2021-12-07-600},
  title = {Physical {I}mplementability of {L}inear {M}aps and {I}ts {A}pplication in {E}rror {M}itigation},
  author = {Jiang, Jiaqing and Wang, Kun and Wang, Xin},
  journal = {{Quantum}},
  issn = {2521-327X},
  publisher = {{Verein zur F{\"{o}}rderung des Open Access Publizierens in den Quantenwissenschaften}},
  volume = {5},
  pages = {600},
  month = dec,
  year = {2021}
}

@article{PhysRevResearch.3.033178,
  title = {Optimal resource cost for error mitigation},
  author = {Takagi, Ryuji},
  journal = {Phys. Rev. Res.},
  volume = {3},
  issue = {3},
  pages = {033178},
  numpages = {11},
  year = {2021},
  month = {Aug},
  publisher = {American Physical Society},
  doi = {10.1103/PhysRevResearch.3.033178},
  url = {https://link.aps.org/doi/10.1103/PhysRevResearch.3.033178}
}

@article{PhysRevA.109.022410,
  title = {Efficacy of virtual purification-based error mitigation on quantum metrology},
  author = {Kwon, Hyukgun and Oh, Changhun and Lim, Youngrong and Jeong, Hyunseok and Jiang, Liang},
  journal = {Phys. Rev. A},
  volume = {109},
  issue = {2},
  pages = {022410},
  numpages = {12},
  year = {2024},
  month = {Feb},
  publisher = {American Physical Society},
  doi = {10.1103/PhysRevA.109.022410},
  url = {https://link.aps.org/doi/10.1103/PhysRevA.109.022410}
}

@article{liuVirtualChannelPurification2024,
  title = {Virtual Channel Purification},
  author = {Liu, Zhenhuan and Zhang, Xingjian and Fei, Yue-Yang and Cai, Zhenyu},
  journal = {PRX Quantum},
  volume = {6},
  issue = {2},
  pages = {020325},
  numpages = {33},
  year = {2025},
  month = {May},
  publisher = {American Physical Society},
  doi = {10.1103/PRXQuantum.6.020325},
  url = {https://link.aps.org/doi/10.1103/PRXQuantum.6.020325}
}

@article{zcai_paulitwirling19,
	author = {Cai, Zhenyu and Benjamin, Simon C.},
	date = {2019/08/02},
	doi = {10.1038/s41598-019-46722-7},
	id = {Cai2019},
	isbn = {2045-2322},
	journal = {Scientific Reports},
	number = {1},
	pages = {11281},
	title = {Constructing Smaller Pauli Twirling Sets for Arbitrary Error Channels},
	url = {https://doi.org/10.1038/s41598-019-46722-7},
	volume = {9},
	year = {2019}}

@article{PhysRevA.94.052325,
  title = {Noise tailoring for scalable quantum computation via randomized compiling},
  author = {Wallman, Joel J. and Emerson, Joseph},
  journal = {Phys. Rev. A},
  volume = {94},
  issue = {5},
  pages = {052325},
  numpages = {9},
  year = {2016},
  month = {Nov},
  publisher = {American Physical Society},
  doi = {10.1103/PhysRevA.94.052325},
  url = {https://link.aps.org/doi/10.1103/PhysRevA.94.052325}
}

@article{PhysRevLett.117.160801,
  title = {Sequential Feedback Scheme Outperforms the Parallel Scheme for Hamiltonian Parameter Estimation},
  author = {Yuan, Haidong},
  journal = {Phys. Rev. Lett.},
  volume = {117},
  issue = {16},
  pages = {160801},
  numpages = {6},
  year = {2016},
  month = {Oct},
  publisher = {American Physical Society},
  doi = {10.1103/PhysRevLett.117.160801},
  url = {https://link.aps.org/doi/10.1103/PhysRevLett.117.160801}
}

@article{380f7170-a649-307c-9495-f3b3298846ff,
 ISSN = {02643952},
 URL = {http://www.jstor.org/stable/91208},
 author = {R. A. Fisher},
 journal = {Philosophical Transactions of the Royal Society of London. Series A, Containing Papers of a Mathematical or Physical Character},
 pages = {309--368},
 publisher = {Royal Society},
 title = {On the Mathematical Foundations of Theoretical Statistics},
 urldate = {2025-02-26},
 volume = {222},
 year = {1922}
}

@article{PhysRevA.53.2855,
  title = {Five two-bit quantum gates are sufficient to implement the quantum Fredkin gate},
  author = {Smolin, John A. and DiVincenzo, David P.},
  journal = {Phys. Rev. A},
  volume = {53},
  issue = {4},
  pages = {2855--2856},
  numpages = {0},
  year = {1996},
  month = {Apr},
  publisher = {American Physical Society},
  doi = {10.1103/PhysRevA.53.2855},
  url = {https://link.aps.org/doi/10.1103/PhysRevA.53.2855}
}

@article{chuang1997prescription,
  title={Prescription for experimental determination of the dynamics of a quantum black box},
  author={Chuang, Isaac L and Nielsen, Michael A},
  journal={Journal of Modern Optics},
  volume={44},
  number={11-12},
  pages={2455--2467},
  year={1997},
  publisher={Taylor \& Francis}
}

@article{PhysRevA.87.062119,
  title = {Self-consistent quantum process tomography},
  author = {Merkel, Seth T. and Gambetta, Jay M. and Smolin, John A. and Poletto, Stefano and C\'orcoles, Antonio D. and Johnson, Blake R. and Ryan, Colm A. and Steffen, Matthias},
  journal = {Phys. Rev. A},
  volume = {87},
  issue = {6},
  pages = {062119},
  numpages = {9},
  year = {2013},
  month = {Jun},
  publisher = {American Physical Society},
  doi = {10.1103/PhysRevA.87.062119},
  url = {https://link.aps.org/doi/10.1103/PhysRevA.87.062119}
}

@article{PhysRevLett.131.210602,
  title = {Universal Sampling Lower Bounds for Quantum Error Mitigation},
  author = {Takagi, Ryuji and Tajima, Hiroyasu and Gu, Mile},
  journal = {Phys. Rev. Lett.},
  volume = {131},
  issue = {21},
  pages = {210602},
  numpages = {8},
  year = {2023},
  month = {Nov},
  publisher = {American Physical Society},
  doi = {10.1103/PhysRevLett.131.210602},
  url = {https://link.aps.org/doi/10.1103/PhysRevLett.131.210602}
}

@article{PhysRevLett.131.210601,
  title = {Universal Cost Bound of Quantum Error Mitigation Based on Quantum Estimation Theory},
  author = {Tsubouchi, Kento and Sagawa, Takahiro and Yoshioka, Nobuyuki},
  journal = {Phys. Rev. Lett.},
  volume = {131},
  issue = {21},
  pages = {210601},
  numpages = {7},
  year = {2023},
  month = {Nov},
  publisher = {American Physical Society},
  doi = {10.1103/PhysRevLett.131.210601},
  url = {https://link.aps.org/doi/10.1103/PhysRevLett.131.210601}
}

@article{endo22,
	author = {Takagi, Ryuji and Endo, Suguru and Minagawa, Shintaro and Gu, Mile},
	date = {2022/09/22},
	doi = {10.1038/s41534-022-00618-z},
	id = {Takagi2022},
	isbn = {2056-6387},
	journal = {npj Quantum Information},
	number = {1},
	pages = {114},
	title = {Fundamental limits of quantum error mitigation},
	url = {https://doi.org/10.1038/s41534-022-00618-z},
	volume = {8},
	year = {2022}}

@misc{cai2023practicalframeworkquantumerror,
      title={A Practical Framework for Quantum Error Mitigation}, 
      author={Zhenyu Cai},
      year={2023},
      eprint={2110.05389},
      archivePrefix={arXiv},
      primaryClass={quant-ph},
      url={https://arxiv.org/abs/2110.05389}, 
}

@misc{kwon2025virtualpurificationcomplementsquantum,
      title={Virtual purification complements quantum error correction in quantum metrology}, 
      author={Hyukgun Kwon and Changhun Oh and Youngrong Lim and Hyunseok Jeong and Seung-Woo Lee and Liang Jiang},
      year={2025},
      eprint={2503.12614},
      archivePrefix={arXiv},
      primaryClass={quant-ph},
      url={https://arxiv.org/abs/2503.12614}, 
}

@article{roy2009unitary,
  title={Unitary designs and codes},
  author={Roy, Aidan and Scott, Andrew J},
  journal={Designs, codes and cryptography},
  volume={53},
  pages={13--31},
  year={2009},
  publisher={Springer}
}

@article{zhou18,
	author = {Zhou, Sisi and Zhang, Mengzhen and Preskill, John and Jiang, Liang},
	date = {2018/01/08},
	doi = {10.1038/s41467-017-02510-3},
	id = {Zhou2018},
	isbn = {2041-1723},
	journal = {Nature Communications},
	number = {1},
	pages = {78},
	title = {Achieving the Heisenberg limit in quantum metrology using quantum error correction},
	url = {https://doi.org/10.1038/s41467-017-02510-3},
	volume = {9},
	year = {2018}
}

@article{Shettell_2021,
doi = {10.1088/1367-2630/abf533},
url = {https://dx.doi.org/10.1088/1367-2630/abf533},
year = {2021},
month = {apr},
publisher = {IOP Publishing},
volume = {23},
number = {4},
pages = {043038},
author = {Shettell, Nathan and Munro, William J and Markham, Damian and Nemoto, Kae},
title = {Practical limits of error correction for quantum metrology},
journal = {New Journal of Physics}
}

\end{document}